%% file: Paper_Jan22_2014-unblinded.tex
\documentclass[notitlepage,12pt]{article}
\usepackage{graphicx,bbm}
\usepackage{amsfonts}
\usepackage{amsmath}
\usepackage[twoside]{geometry}
\usepackage[small]{caption2}
\usepackage[onehalfspacing]{setspace}
\usepackage{scalefnt}
\usepackage{natbib}
\usepackage{natbib}
\usepackage{natbib}

\newenvironment{my_enumerate}{
\begin{enumerate}
  \setlength{\itemsep}{1pt}
  \setlength{\parskip}{0pt}
  \setlength{\parsep}{0pt}}{\end{enumerate}
}

\begin{document}

\title{Bayesian modeling and forecasting of 24-hour high- frequency volatility: A case study of the financial crisis}
\author{Jonathan Stroud and Michael Johannes\thanks{%
Jonathan Stroud is Associate Professor, Department of Statistics, George Washington University
(stroud@gwu.edu). Michael Johannes is Professor, Finance and Economics Division, Columbia 
Business School (mj335@columbia.edu). }$^{\ast }$ \\
George Washington University and Columbia Business School}
\maketitle

\begin{abstract}
\setcounter{page}{0} \thispagestyle{empty} This paper estimates models of
high frequency index futures returns using `around the clock' 5-minute
returns that incorporate the following key features:\ multiple persistent
stochastic volatility factors, jumps in prices and volatilities, seasonal
components capturing time of the day patterns, correlations between return
and volatility shocks, and announcement effects. We develop an integrated
MCMC\ approach to estimate interday and intraday parameters and states using
high-frequency data without resorting to various aggregation measures like
realized volatility. We provide a case study using financial crisis data
from 2007 to 2009, and use particle filters to construct likelihood
functions for model comparison and out-of-sample forecasting from 2009 to
2012. We show that our approach improves realized volatility forecasts by up
to 50\% over existing benchmarks.


\newpage
\end{abstract}


\section{Introduction}

Financial crises are a rich information source to learn about asset price
dynamics and models used to capture these dynamics. For example, the 1987
Crash and 1998 LTCM hedge fund crisis highlighted the importance of
stochastic volatility (SV) and jumps, in both prices and volatility, for
understanding index returns 
\cite*[see, e.g.,][]{Bate:00,DuffPanSing:00,ErakJohaPols:03,Todo:11}
The recent crisis provides similar opportunities largely due to two unique 
features. First, unlike the 1987 or 1998 crises which were short-lived, the 
recent crisis began in mid 2007 and
lasted well into 2009, with aftershocks into the European debt crisis and
Flash-Crash in 2010. Second, structural changes in the mid 2000s led to
continuous around-the-clock markets, as markets migrated from traditional
floor execution during `regular' market hours to fully electronic 24-hour
trading.  For the first time, there is `around the clock' high frequency 
data in a long-lasting crisis.

This paper uses newly available data to study a range of important
questions. What sort of models and factors are required to accurately model
24-hour high-frequency crisis returns? Do these specifications generate
dynamics similar to extant ones?\ How useful are these models for practical
applications like return distribution and volatility forecasting or trading?
Answers to these questions are important for academics, policy makers,
market participants and risk managers who need to understand the structure
of financial market volatility and to quantify the likelihood of potential
future market movements. In particular, nearly every practical finance
application -- including optimal investments and trading,
options/derivatives pricing, market making and market microstructure, and
risk management -- requires volatility forecasts.

\begin{figure}[tbp]
\begin{center}
\includegraphics[height=3.5in,width=5.5in]{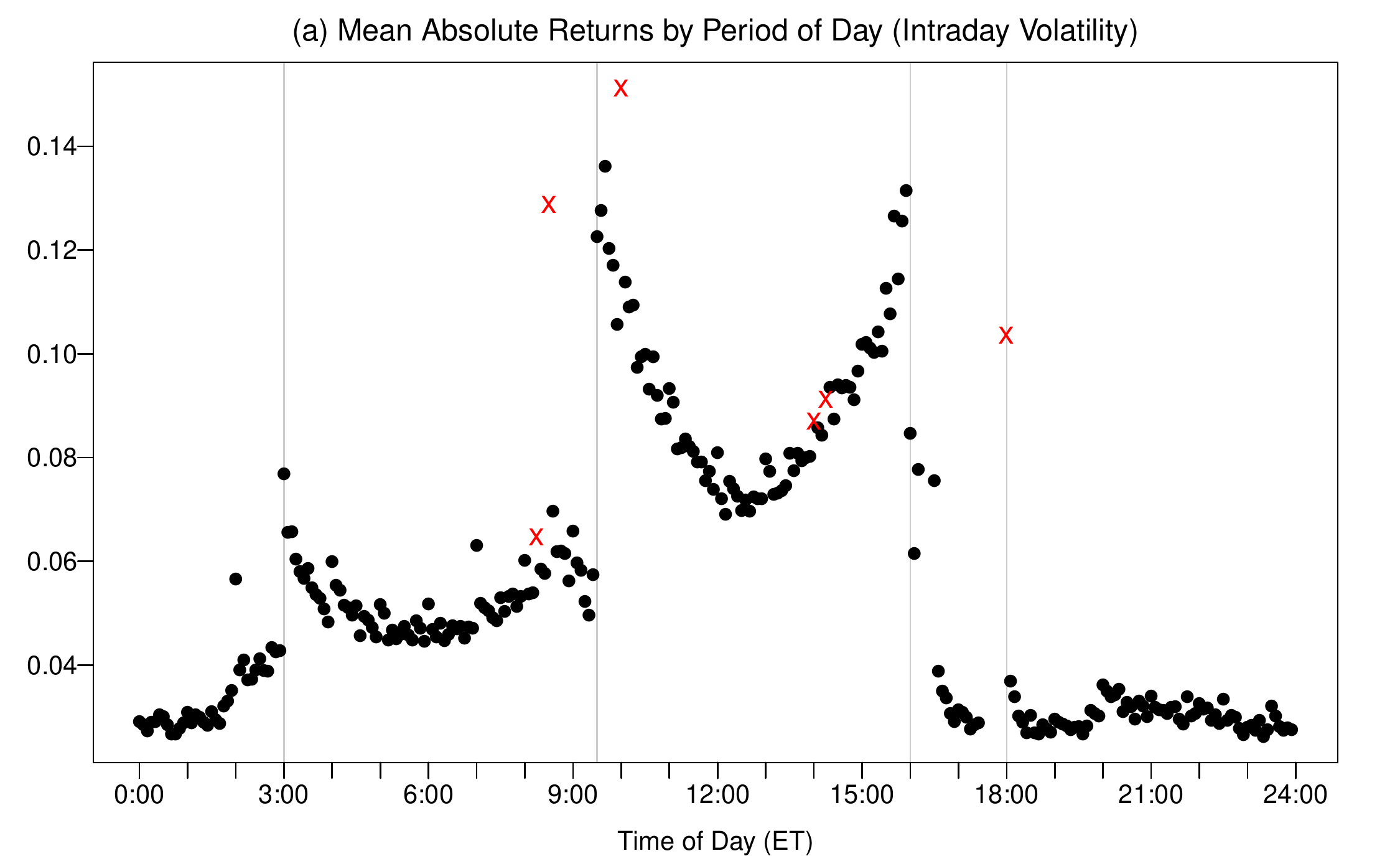}
\par
\vspace{.3cm} %
\includegraphics[height=3.5in,width=5.5in]{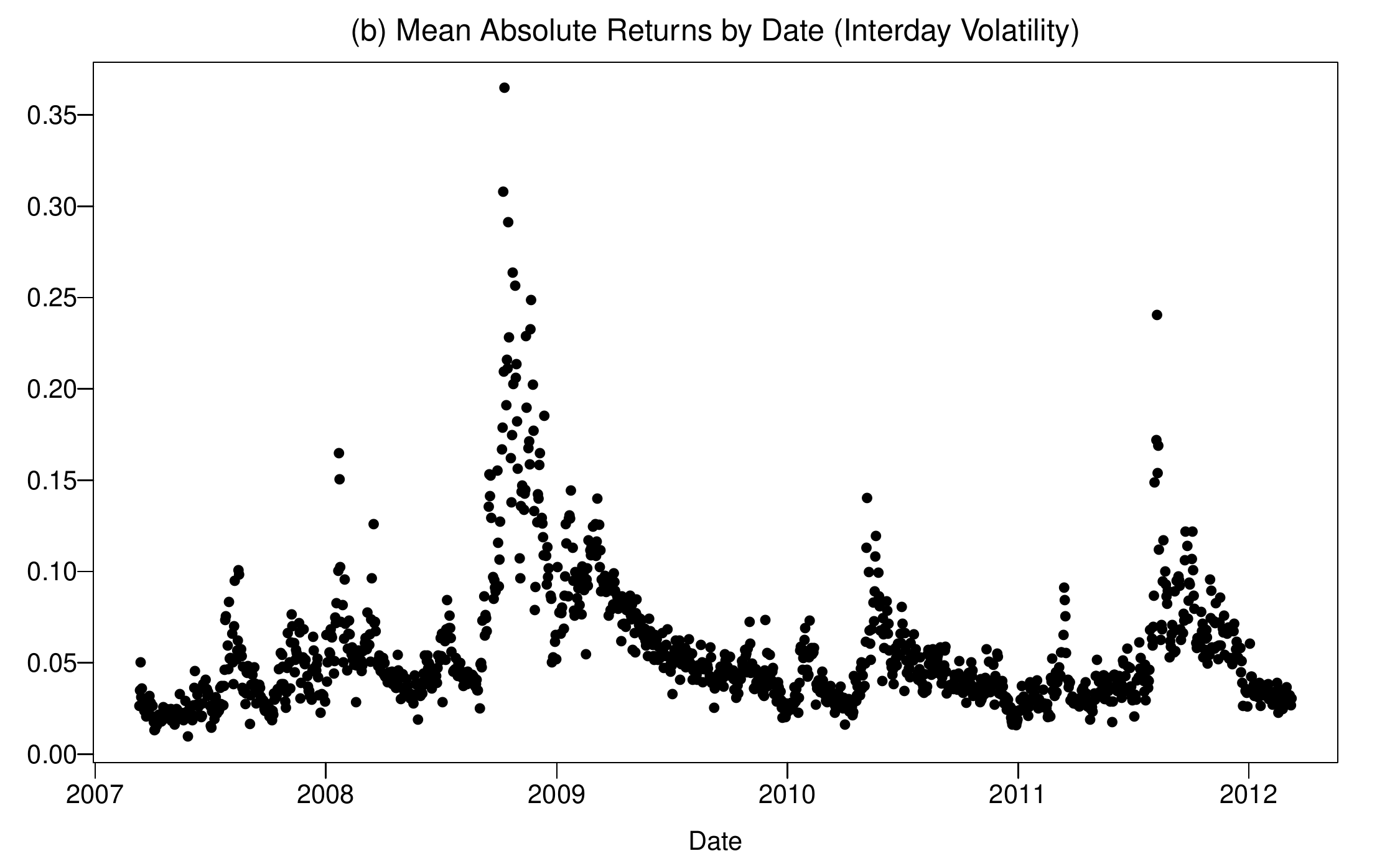}
\end{center}
\caption{Summary of five-minute returns on S\&P E-mini futures prices, March
2007 -- March 2012. (a) Mean absolute returns for each period of the day.
The trading day runs from 18:00 ET--17:30 ET, with a break in trading from
16:15--16:30. Macroeconomic announcement times are marked with an 'x', and
selected major market open and closing times are marked with vertical lines.
(b) Mean absolute returns by date.}
\label{Period_vol}
\end{figure}

Our case study focuses on the S\&P 500 index, arguably the world's most
important asset market, using S\&P 500 index futures, which trade 24 hours 
a day from Sunday evening to Friday night. We focus on in-sample model
fitting, which allows us to learn about the underlying structure of returns,
and fully out-of-sample prediction, which is important for applications. 
We use parametric models estimated from intraday returns, something rarely 
attempted due to data complexities and computational burdens. Figure 
\ref{Period_vol} plots intraday and interday volatility of 5-minute S\&P 500 
futures returns from March 2007 to March 2012.  Intraday volatility has 
complicated, periodic patterns driven by the global migration of trading and 
macroeconomic announcements \cite[see e.g.][]{AndeBoll:97,AndeBoll:98}.
Interday volatility is persistent, stochastic, and
mean-reverting. Models capturing these components require multiple
volatility factors, complicated shocks, and many parameters, which, in
conjunction with huge volumes of high-frequency data, make parametric
estimation difficult.

Due to these complexities, most researchers use nonparametric `realized
volatility' (RV) methods to avoid directly modeling intraday returns by
aggregating intraday data into a daily RV measure 
\citep[see][for reviews]{AndeBenz:09,BarnShep:07}. One weakness is
its nonparametric nature: RV approaches generally do not specify a full
model of returns, which limits practical usefulness as there is no return
distribution, just volatility estimates. Despite this weakness, RV methods
are extremely useful and are a popular volatility forecasting approach.

Methodologically, we build new models with the flexibility to fit the
complexities of 24-hour intraday data during the financial crisis. We 
develop novel MCMC algorighms to fit models in-sample and use particle 
filters to compute predictive distributions and volatility forecasts for 
out-of-sample validation. Although SV models are commonly implemented with 
MCMC, we know of no applications using realistic SV models and intraday 
data for out-of-sample validation.

We find strong in and out-of-sample evidence for multiscale volatility with
distinct `fast' and `slow' factors. The slow factor's half-life is about 25
days, similar to extant estimates from daily data. The fast factor, however,
operates intradaily, with a half-life of an hour, capturing the `digestion'
time of high-frequency news or liquidity events. Our models offer a
significant improvement over traditional GARCH models estimated on intraday
data. We find strong evidence for jumps (in prices and volatility) or
fat-tails generated by t-distributed return shocks. Price jumps are rather
small in comparison to estimates from earlier periods or option prices which
identify jumps as large negative `crashes.' This could be unique to the
recent crisis or something more fundamental uncovered from newly available
high frequency data. A striking and important features of our analysis is a
strong and uniform ranking of models both in and out-of-sample based on
predictive likelihoods.

The ultimate test of a model is usefulness, and we consider three practical
applications: volatility forecasting, tail risk management, and a trading
application. We compare our models' performance to popular GARCH and RV
benchmarks. In forecasting volatility, our SV models generate significantly
lower forecasting errors than all competitors at all horizons. The absolute
performance is striking as we generate fully out-of-sample volatility
forecasts with R%
${{}^2}$%
's in excess of 70\%. Our SV models perform relatively and absolutely well
in a quantitative risk management application--evaluating the accuracy of
value-at-risk (VaR) forecasts, essentially tail forecasting--and a simple
volatility trading application. Overall, we find strong evidence for the
usefulness of our models and approach in all cases.

\section{Data, modeling and estimation approach}

\subsection{Data}

This paper studies S\&P 500 index futures. Two contract variants exist:\ the
traditional `full-size' contract (\$250 per index point) and the `E-mini'
contract (\$50 per point). E-minis trade electronically on the Chicago
Mercantile Exchange's (CME's) Globex platform and initially complemented the
full-sized contract, which traded in a traditional `open outcry' pit. E-mini
trading volumes increased steadily before expanding rapidly in 2007 
\citep*[see CME Group,][]{LabuNyhoCo:10} with the advent of algorithmic 
high-frequency trading and increased global influences.  S\&P 500 futures are 
one of the most liquid contracts in the world, limiting any microstructure 
effects \citep*[see, e.g.,][]{CorsMittPigo:08}.

We analyze 5-minute data from March 11, 2007 through March 9, 2012, consisting 
of 352,887 5-minute observations over 1293 days. The price data is for
quarterly contracts, which are converted to a `continuous contract' by
rolling contracts two weeks before expiration. The first two years are used
for parameter estimation and the remaining for forecasting. March 2007 is
a natural starting date as it coincides with the dramatic trading volume
increase. Trading starts on Sunday night at 18:00 and continues until 16:15
Friday (all times are Eastern). Markets close Monday-Thursday from
16:15-16:30 and from 17:30-18:00. Sunday `open' returns are from Friday at
16:15 to Sunday at 18:00. There are similar `open' returns from 16:15-16:30
and 17:30-18:00. Normal days have 279 return observations.

\subsection{Stochastic volatility models}

We model 5-minute logarithmic price returns, $y_{t}$, which evolve via 
\begin{equation*}
y_{t}=100\cdot \log \left( P_{t}/P_{t-1}\right) =\mu +v_{t}\varepsilon
_{t}^{\ast }+J_{t}Z_{t}^{y}\text{,}
\end{equation*}%
where $P_{t}$ is the futures price, $\mu $ is the mean return, $v_{t}$ is
diffusive or non-jump volatility, $J_{t}$ is a jump indicator with $P\left[%
J_{t}=1\right] =\kappa $, $Z_{t}^{y}\overset{i.i.d.}{\sim }\mathcal{N}(\mu
_{y},\sigma _{y}^{2})$ are return jumps, and $\varepsilon _{t}^{\ast }=\sqrt{%
\lambda _{t}}\varepsilon _{t}$ where $\varepsilon _{t}\overset{i.i.d.}{\sim }%
\mathcal{N}(0,1)$ and $\lambda _{t}\overset{i.i.d.}{\sim }\mathcal{IG}\left(
\nu /2,\nu /2\right) $, which implies $\varepsilon _{t}^{\ast }\overset{%
i.i.d.}{\sim }t_{v}\left( 0,1\right) $. At this level, the model resembles
common jump-diffusion specifications.

There is strong evidence for stochastic volatility and jumps in S\&P 500
index prices from daily data \cite[e.g.,][]{ErakJohaPols:03}, index option 
prices \citep*{BaksCaoChen:97,Bate:00,DuffPanSing:00}, and intraday data 
\cite[][provide a review]{AndeShep:09}.  Estimates from options or daily 
returns identify large jumps or `crashes.' Studies using recent high 
frequency data tend to find smaller jumps, though these studies typical 
ignore overnight periods.

We model total volatility via a multiplicative specification: 
\begin{equation}
v_{t}=\sigma \cdot X_{t,1}\cdot X_{t,2}\cdot S_{t}\cdot A_{t}\text{,}
\end{equation}%
where $X_{t,1}$ and $X_{t,2}$ are SV processes, and $S_{t}$/$A_{t}$ are
seasonal/announcement components. $\sigma $ is interpreted as the modal
volatility (i.e. $v_{t}$ when $X_{t,1}=X_{t,2}=S_{t}=A_{t}=1$). 
The log of total diffusive variance is linear:
\begin{equation}
h_{t}=\log (v_{t}^{2})=\mu _{h}+x_{t,1}+x_{t,2}+s_{t}+a_{t},  \label{logvol}
\end{equation}%
where $\mu _{h}=\log (\sigma ^{2})$, $x_{t,i}=\log (X_{t,i}^{2}),s_{t}=\log
(S_{t}^{2})$, and $a_{t}=\log (A_{t}^{2})$.

Volatility evolves stochastically via 
\begin{equation*}
x_{t+1,1}=\phi_{1}x_{t,1}+\sigma _{1}\eta _{t,1}\text{  and  }
x_{t+1,2}=\phi_{2}x_{t,2}+\sigma _{2}\eta _{t,2}+J_{t}Z_{t}^{v},
\end{equation*}%
where $\eta _{t,i}\overset{i.i.d.}{\sim }\mathcal{N}(0,1)$ and $Z_{t}^{v}%
\overset{i.i.d.}{\sim }\mathcal{N}(\mu _{v},\sigma _{v}^{2})$ are the jumps
in log-volatility. Notice the volatility jump times are coincident with those 
in returns. $\rho =corr(\varepsilon _{t},\eta _{t,2})$ captures diffusive
\textquotedblleft leverage\textquotedblright\ effects via correlated shocks
to returns and fast volatility. We assume a multiscale volatility
specification, assuming $0<\phi _{2}<\phi _{1}<1$, with $X_{t,1}$ and $%
X_{t,2}$ the `slow' and `fast' volatility factors, respectively. Both
factors are affected by intraday shocks, relaxing a common assumption that
stochastic volatility is constant intraday 
\citep[see, e.g.,][]{AndeBoll:97,AndeBoll:98}.

We model the seasonal/periodic and announcement effects as deterministic
volatility patterns using the spline approach in \cite*{WeinBrowStro:07}.
The seasonal component is $s_{t}=f_{t}^{\prime }\beta $, 
where $f_{t}=(f_{t1},...,f_{t,288})^{\prime }$ is an indicator vector 
where $f_{tk}=1$ if time $t$ occurs at period of the day $k$ and zero otherwise,
and $\beta =(\beta _{1},...,\beta _{288})^{\prime }$ are the seasonal
coefficients.  We impose the constraint $\sum\nolimits_{k=1}^{288}\beta_k=0$ for 
identification.  To incorporate smoothness in the seasonal coefficients
we assume a cubic smoothing spline prior for $\beta$, with discontinuities 
at market opening/closing times.  Following \cite{Wahb:78} and \cite{KohnAnsl:87}, 
we write this as a multivariate normal prior $\beta \sim \mathcal{N}(0,\tau _{s}^{2}U_{s})$,
where $\tau _{s}^{2}$ is the smoothing parameter and $U_{s}$ 
is a known correlation matrix (see Appendix C).

The announcement component is $a_{t}=\sum_{i=1}^{n}I_{ti}^{\prime }\alpha
_{i}$, where $I_{ti}=(I_{ti1},...,I_{ti5})^{\prime }$ is an indicator
vector for news type $i$ with $I_{tik}=1$ if a news release occurred at period $%
t-k$ and zero otherwise, and $\alpha_{i}=(\alpha _{i1},...,\alpha_{i5})^{\prime}$ 
are the announcement effects for news type $i$. We again
assume cubic smoothing spline priors to smooth the coefficients,
$\alpha _{i}\sim \mathcal{N}(0,\tau
_{a}^{2}U_{a})$ (see Appendix C). We consider $n=14$ announcement types
listed in Table~\ref{announcements} in the Appendix. We assume that 
announcements increase market volatility for $K=5$ periods, i.e., markets digest 
the news in 25 minutes. Sunday open is treated as an announcement.

Our model applies to all 5-minute intraday returns, not just those
`traditional' trading hours from 9:30 to 16:00. Existing papers often either
ignore or simplistically correct for overnight returns. For example, 
\cite{EnglSoka:12}, following \textquotedblleft common
practice,\textquotedblright\ delete overnight returns due either to a lack
of overnight data (for individual stocks) or difficulties in modeling
overnight returns, which requires both periodic and announcement components.
Ignoring overnight returns is problematic for 24-hour, global markets and
crisis periods. For example, on October 24, 2008, S\&P 500 futures fell over 6
percent overnight, and deleting this period would remove important
information.

\subsection{Estimation approach}

We take a Bayesian approach and use MCMC to simulate from the posterior
distribution, 
\begin{equation*}
p\left( z^{T},\beta ,\alpha ,\theta ,|y^{T}\right) \,\propto
\prod_{t=1}^{T}\,p(y_{t}|z_{t},\beta ,\alpha ,\theta
)\,p(z_{t}|z_{t-1},\theta )\,p(\beta |\theta )\,p(\alpha |\theta )\,p(\theta
),
\end{equation*}%
where $z_{t}=\left( x_{t,1},x_{t,2},\lambda
_{t},J_{t},Z_{t}^{y},Z_{t}^{v}\right) $, $z^{T}=\left(
z_{1},...,z_{T}\right) $, $\theta $ are parameters and $y^{T}=\left(
y_{1},...,y_{T}\right) $ are returns. Appendices A and D detail the priors
and algorithm, respectively. We use standard conjugate priors where possible
and in all cases proper, though not strongly informative, priors.
Efficiently programmed in C, the MCMC algorithm makes 12,500 draws in 12--25
minutes using a 2.8 GHz Xeon processor for each year of 5-minute returns
(around 70,500 observations). Computing time is approximately linear for the
sample sizes considered.

Our algorithm is highly tuned using representation and sampling `tricks.' We
express the model as a linear, but non-Gaussian system and use the 
\cite{CartKohn:94} and \cite{Fruh:94} forward-filtering,
backward sampling algorithm for block updating, an approach first used for
SV models in \cite*{KimShepChib:98}.  When possible, parameters and
states are drawn together. Following \cite{AnslKohn:87}, we express the
splines as a state space model and update in blocks.  Building on the 
methodology of \cite*{JohaPolsStro:09}, we use auxiliary particle filters 
\citep{PittShep:99} to approximately sample from 
$p\left(z_{t}|y^{t},\widehat{\theta }\right)$, where $\widehat{\theta }$ 
is the posterior mean. Appendix E provides details.

It is useful to contrast our intraday parametric estimation approach to
\cite{AndeBoll:97,AndeBoll:98}, the main competitor. They model
5-minute exchange rates via long-memory GARCH models with seasonal effects 
\citep*[see also][]{MartChanTayl:02} and use a
two-step procedure to first estimate daily volatility, assumed constant
intraday, and then estimate a flexible seasonal component. 
\cite{EnglSoka:12} estimate GARCH models on intraday returns for 2500
individual stocks with a seasonal component using third-party interday
volatility estimates. By contrast, we simultaneously estimate all parameters
and states, avoiding the need for potentially inefficient two-stage
estimators and restrictive assumptions like normally distributed shocks and
the absence of jumps.

Another approach aggregates intraday returns into daily RV statistics, which
are used to estimate models at a daily frequency 
\citep[see, e.g.,][]{BarnShep:02,Todo:11}.  We estimate the
models directly on 5-minute returns, without aggregation into RV, which
allows us to identify intraday components and forecast at high frequencies.
\cite{HansHuanShek:12} introduce a hybrid model, called Realized GARCH 
(RealGARCH), combining the tractability of daily GARCH models with the
information in realized volatility. We implement these promising models and
compare their performance to our SV models.

Appendix D provides algorithm details, with diagnostics in the web Appendix.
The MCMC algorithms mix quite well given the large number of unknown states
and parameters, although models with jumps in volatility mix more slowly
than those with only diffusive volatility, and volatility of volatility
parameters mix relatively slowly. Parameters deep in the state space (e.g.,
volatility of volatility) tend to traverse the state space more slowly,
consistent with multiple layers of smoothing \cite[see, e.g.,][]{KimShepChib:98}.
This does not mean that these parameters are not accurately
estimated, as simulation evidence does indicate they can be accurately
estimated. The only model with any substantive concern is the SVCJ$_{2}$
model, and we thin the samples to alleviate any concerns.  We have also
considered significantly less informative priors and the results do not
substantially change.

\subsection{Decompositions and Diagnostics}

To decompose variance and to quantify relative importance, we compute the
posterior mean for the total log variance and for each variance component at
each time period, e.g., $\overline{x}_{t,1}=E\left[ x_{t,1}|y^{T}\right] $,
run univariate regressions of the form $\overline{h_{t}}=\alpha +\beta 
\overline{x}_{t,1}+\varepsilon _{t}$, and report $R^{2}$'s for each
component. We report decompositions in both log-variance and in volatility
units.

To quantify model fit, we would ideally use the Bayes factor, $\mathcal{B}%
_{i,j}^{t}=\mathbb{P}\left[ \mathcal{M}_{i}|y^{t}\right] /\mathbb{P}\left[ 
\mathcal{M}_{j}|y^{t}\right] ,$ where $\left\{ \mathcal{M}_{i}\right\}
_{i=1}^{M}$ indicate models, $\mathbb{P}\left[ \mathcal{M}_{i}|y^{t}\right]
\propto p\left( y^{t}|\mathcal{M}_{i}\right) \mathbb{P}\left( \mathcal{M}%
_{i}\right) $, and $p\left( y^{t}|\mathcal{M}_{i}\right) $ is the marginal
likelihood. Bayes factors are often called an \textquotedblleft automated
Occam's razor,\textquotedblright\ as they penalize loosely parameterized
models \citep{SmitSpie:80}. Computing marginal likelihoods
requires sequential parameter estimation, which is computationally
prohibitive, so we alternatively report log-likelihoods and the Bayesian
Information Criterion (BIC) statistic, which approximates the Bayes factor.

The model $\mathcal{M}_{i}$ likelihood of $y^{T}$ is%
\begin{equation*}
\mathcal{L}\left( y^{T}|\theta _{\left( i\right) },\mathcal{M}_{i}\right)
=\prod\nolimits_{t=0}^{T-1}p\left( y_{t+1}|\theta _{\left( i\right) },y^{t},%
\mathcal{M}_{i}\right) ,
\end{equation*}%
where $\theta _{\left( i\right) }$ are the parameters in $\mathcal{M}_{i}$, $%
p\left( y_{t+1}|\theta _{\left( i\right) },y^{t},\mathcal{M}_{i}\right) $ is
the predictive return distribution, 
\begin{equation*}
p\left( y_{t+1}|\theta _{\left( i\right) },y^{t},\mathcal{M}_{i}\right)
=\int p\left( y_{t+1}|\theta _{\left( i\right) },z_{t+1},\mathcal{M}%
_{i}\right) p\left( z_{t+1}|\theta _{\left( i\right) },y^{t},\mathcal{M}%
_{i}\right) dz_{t+1},
\end{equation*}%
$p\left( y_{t+1}|\theta _{\left( i\right) },z_{t+1},\mathcal{M}_{i}\right) $
is the conditional likelihood, and $p\left( z_{t+1}|\theta _{\left( i\right)
},y^{t},\mathcal{M}_{i}\right) $ is the state predictive distribution. Given
approximate samples from $p\left( z_{t}|y^{t},\widehat{\theta }_{\left(
i\right) },\mathcal{M}_{i}\right) $, it is easy to approximately sample from
the predictive distributions and likelihoods. All distributions can be
computed at 5-minute and lower frequencies, such as hourly or daily, via
simulation.

Defining the dimensionality of $\theta _{\left( i\right) }$ as $d_{i}$ in
model $\mathcal{M}_{i}$, the BIC criterion is 
\begin{equation*}
BIC_{T}\left( \mathcal{M}_{i}\right) =-2\log \mathcal{L}\left( y^{T}|%
\widehat{\theta }_{\left( i\right) },\mathcal{M}_{i}\right) +d_{i}\log
\left( T\right) \text{.}
\end{equation*}%
BIC and Bayes factors are related asymptotically $BIC_{T}\left( \mathcal{M}%
_{i}\right) -BIC_{T}\left( \mathcal{M}_{j}\right) \approx -2\log \mathcal{B}%
_{i,j}^{T}$ \citep{KassRaft:95}.  BIC asymptotically (in $T$) approximates
the posterior probability of a given model. The dimensionality or degrees of
freedom are not preset for the splines, but are determined by the degree of
fitted smoothness. We compute the degrees of freedom using the state-space
approach of \cite{AnslKohn:87}, evaluating the degrees of freedom at
each iteration of the MCMC algorithm and using the posterior mean for model
comparisons. Given our sample sizes, this approximation should perform well.

\input model-mnemonics-0.tex

For comparisons, we also estimated benchmark GARCH models including a
GARCH(1,1) model (GARCH), and two models that incorporate asymmetry:
the GJR model \citep*{GlosJagaRunk:93}, and the EGARCH model \citep{Nels:91}, 
each with both normal and t-errors fit as in \cite{AndeBoll:97}.   
Appendix G provides details.

\section{Empirical results}

\subsection{In-sample model fits}

Table \ref{Models} describes the models considered. We estimated
single-factor models, but do not report estimates as the 2-factor models
always performed better in and out-of-sample. Table \ref{Modelfit} reports
in-sample fit statistics including the degrees of freedom, log-likelihoods,
and BIC statistics. To ease comparisons, Table \ref{Modelfit} reports Bayes
factors based on the difference of BIC statistics relative to the $SV_{1}$
model, $-2\log \mathcal{B}_{i,SV_{1}}=BIC_{T}\left( \mathcal{M}_{i}\right)
-BIC_{T}\left( \mathcal{M}_{SV_{1}}\right) $. Better fitting models have
higher likelihoods and lower BIC statistics, quantifying the improvement
over a single-factor SV model.

Degrees of freedom range from 253 to 284. This consists of `static'
parameters $d^{\ast }$ (from 4 to 12) and the spline `parameters,' $d_{s}$
and $d_{a},$ which are less than the number of knot points (279 and 70,
respectively) and determined by the spline's smoothness. More complicated
models sometimes have fewer degrees of freedom than their simpler
counterparts, even though they have more static parameters. The multiscale,
two-factor SV models provide the best in-sample fits and, in all cases, the
BIC and log-likelihood statistics provide the same conclusion, which is not
surprising given the large samples. The best performing models, the SVt$_{2}$
and SVCJ$_{2}$ models, have leverage effects and allow for outliers, via
either jumps or $t-$distributed shocks, which are needed to fit the
fat-tails of intraday returns.

\input model-selection-table-2-2-288-long

The multiscale SV models provide significant improvements in fits compared to 
the GARCH models. In fact, the Bayes factors indicate that a simple 1-factor 
SV model actually outperforms all of the GARCH models, strong evidence
supporting SV. This indicates there is something fundamental about the
random nature of volatility in the SV--the extra shock in the volatility
evolution--that improves the fit, which can be compared to the GARCH\ models
in which the shocks to volatility are completely driven by return shocks.

We can not compare likelihood-based fits to RV\ based models which typically
do not specify an intraday return distribution. We also fit variants
omitting seasonalities and/or announcements, which are not reported to save
space. Both components are significant, though the announcement components
are less important given the relatively small number of announcements per
week.

\begin{figure}[tbp]
\par
\begin{center}
\includegraphics[height=3.5in, width=3.15in]{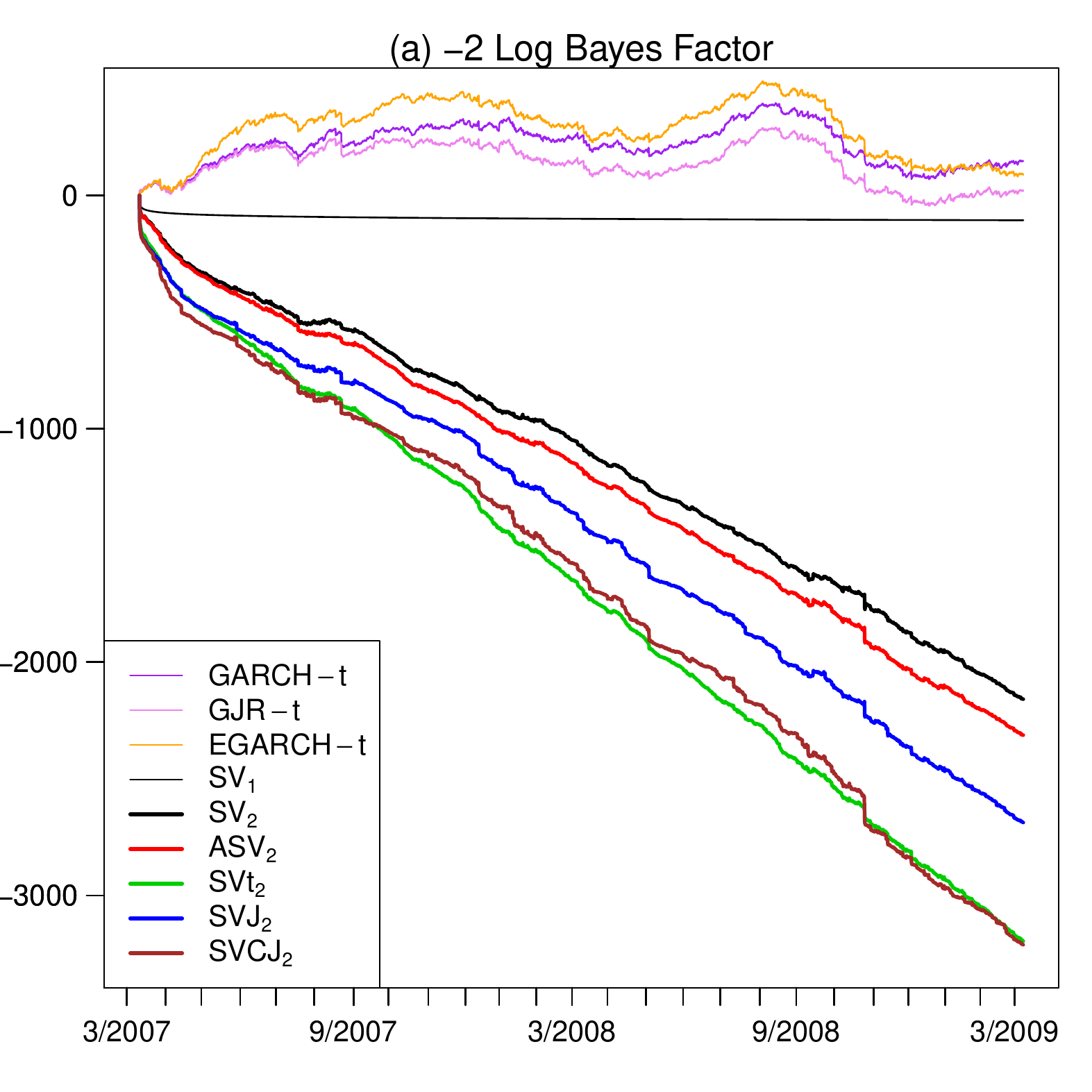}
\includegraphics[height=3.5in, width=3.15in]{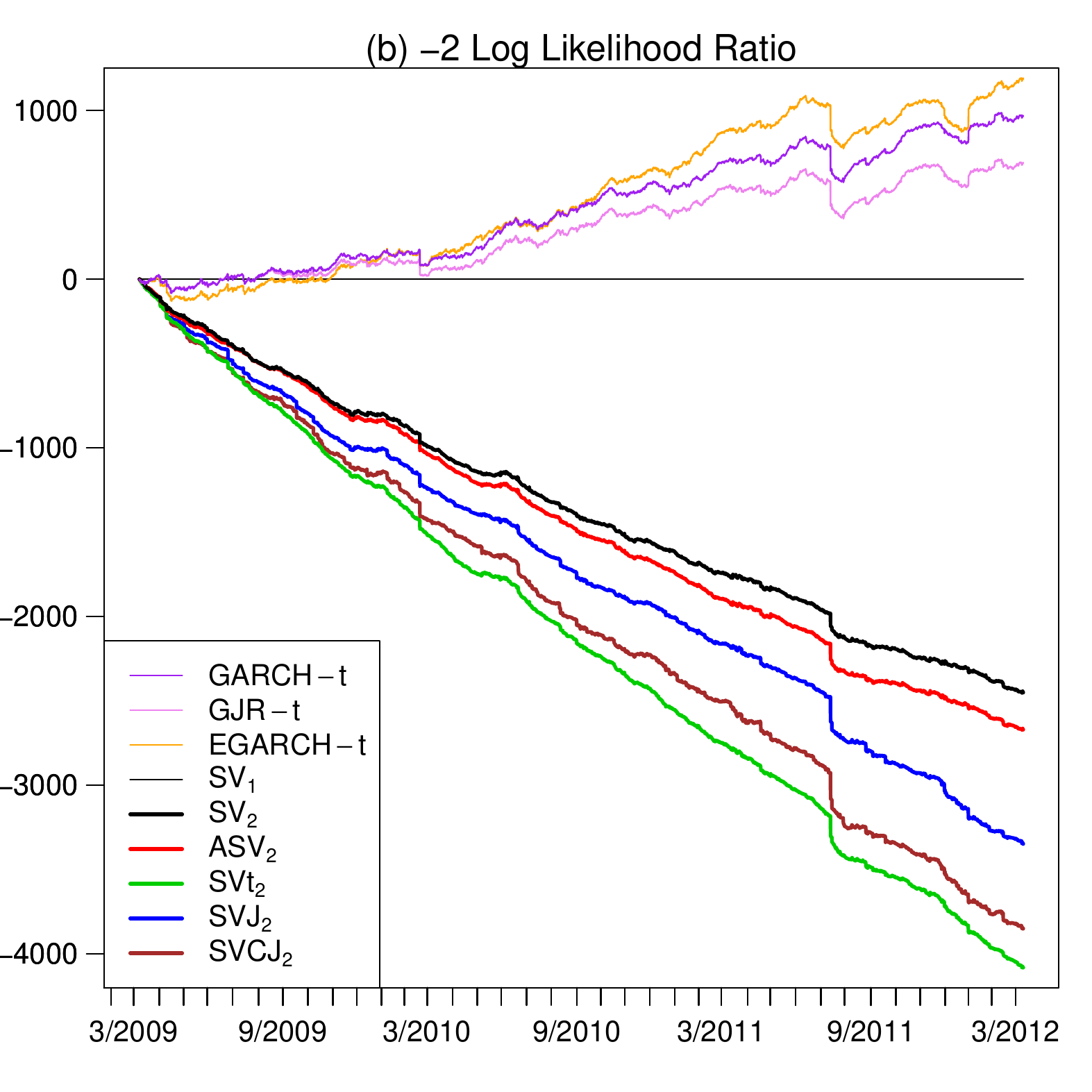}
\end{center}
\caption{(a) Cumulative log Bayes factors during the in-sample period, March
2007--March 2009. (b) Cumulative log-likelihood ratios during the
out-of-sample period, March 2009--March 2012. Values are relative to the SV$%
_1$ model, and are multiplied by -2, so lower values indicate better fit.}
\label{fig:BIC1}
\end{figure}

\cite{West:86} suggests monitoring model fits sequentially through time to
provide an assessment of model failure, either abruptly or slowly over time.
Figure \ref{fig:BIC1}a reports in-sample sequential Bayes factors for each
model relative to the SV$_{1}$ model, $BIC_{T}\left( \mathcal{M}_{i}\right)
-BIC_{T}\left( \mathcal{M}_{SV_{1}}\right) $. Note the gradual
outperformance generated by the SVCJ$_{2}$ and SVt$_{2}$ models, indicating
general fit improvement and not one generated by a very small number of
observations. The relative ranking of the SV models is identical
out-of-sample, confirming the in-sample results.

There is one noticeable spike on October 24, 2008 in the log Bayes factors. This
was caused by a circuit breaker locking S\&P 500 futures limit down from
4:55 am to 9:30 a.m., which generated a number of zero returns. Exchange
rules mandate that S\&P futures can not fall by more than 60 points
overnight and trading can occur at prices above, but not below, this level
until 9:30. Models with fast-moving volatility were able to reduce their
predictive volatility quickly, thus the relatively good fit during this
event. A more complete specification would incorporate a mechanism for limit
down markets.%
%

\subsection{Parameter estimates, variance decompositions and sample paths}

Table~\ref{parameters} summarizes the posteriors and reports inefficiency
factors and acceptance probabilities (for the slowest mixing component, $%
\sigma _{1}$) for the multiscale models. There are a number of interesting
results. The SV factors correspond to a slow-moving interday factor and
rapidly moving intraday factor. Estimates of $\phi _{1}$ in the best fit
models are 0.9999, corresponding to a daily AR(1) coefficient of 0.9725 and
a half-life ($\log 0.5/\log \phi _{1})$ of almost 25 days. This is
consistent with studies using daily data and time-aggregation, that is, that
the data provides similar inference whether sampled at intraday or daily
frequencies. $x_{t,2}$ operates at high frequencies with a 5-minute AR(1)
coefficient $\phi _{2}$ of 0.926 to 0.958, generating a half-life of around
an hour, and high volatility ($\sigma _{2}\gg \sigma _{1}$). Intuitively,
there is strong evidence for rapidly dissipating high-frequency volatility
shocks to volatility. All 2-factor models support an extreme form of
multiscale SV that would be difficult to detect using daily data.

\input param-estimates-two-factor-stacked-288.tex

Decompositions in Table \ref{vardecomps} show the interday factor explains a
majority of total variance, thus the slow-moving factor is relatively more
important than the fast-moving factor. The second factor explains about
7\%-10\% of the total variance. Table \ref{parameters} reports each
volatility factor's unconditional variance, defined as $\tau _{i}^{2}$. $%
\tau _{1}$ is more than twice as large as $\tau _{2}$, driven by the near
unit root behavior of $x_{t,1}\ $and despite $x_{t,1}$'s low conditional
volatility.

The second volatility factor plays a crucial role as it relieves a tension 
present in one-factor models. The SV factor in one-factor models tries to 
fit both low and high-frequency movements, ending up somewhere in between 
and fitting both poorly. For example, in the SVt$_{1}$ model, estimates of 
$\phi_{1}$ are roughly 0.997, corresponding to a daily AR(1)\ coefficient 
of 0.4325 and a half-life of about $0.80$ days, which is much slower than 
the fast factor and much faster than the slow factor in two-factor models. 
The two-factor specifications provide flexibility allowing the factors to 
fit higher and lower frequency volatility fluctuations.

\input vol-decomp-2007-2009-all-1-1-0-2-2-288.tex

Estimates of $\nu $ are about 20, consistent with mild non-normality and
previous daily estimates \citep*[e.g.,][]{ChibNardShep:02,JacqPolsRoss:04}.
Though modest, $\nu $ implies vastly higher probabilities of large shocks, 
some of which will occur in our massive sample. Estimates of $\rho $ are 
modest and around -.10. Identifying this parameter using RV is difficult 
due to various biases \citep[see, e.g.,][]{AitFanLi:13}. Time-variation 
in the variance components accounts for most of the non-normality in models 
without jumps.  Mean jump sizes, $\mu _{y},$ are close to zero in the SVCJ$_{2}$
specification, and arrivals are frequent with $\kappa =.004$ corresponding
to at a rate of 1.17 per day. Return jumps are relatively large as $\sigma _{y}$
is much larger than the modal (non-jump) volatility, e.g., $\sigma
_{y}=0.202 $ vs. $\sigma =0.059$ in the SVCJ$_{2}$ model. Volatility jumps
are quite large, with $\mu _{v}=0.816$ implying that jumps more than double
5-minute volatility.

\begin{figure}[t]
\par
\begin{center}
\includegraphics[height=4in,width=6.2in]{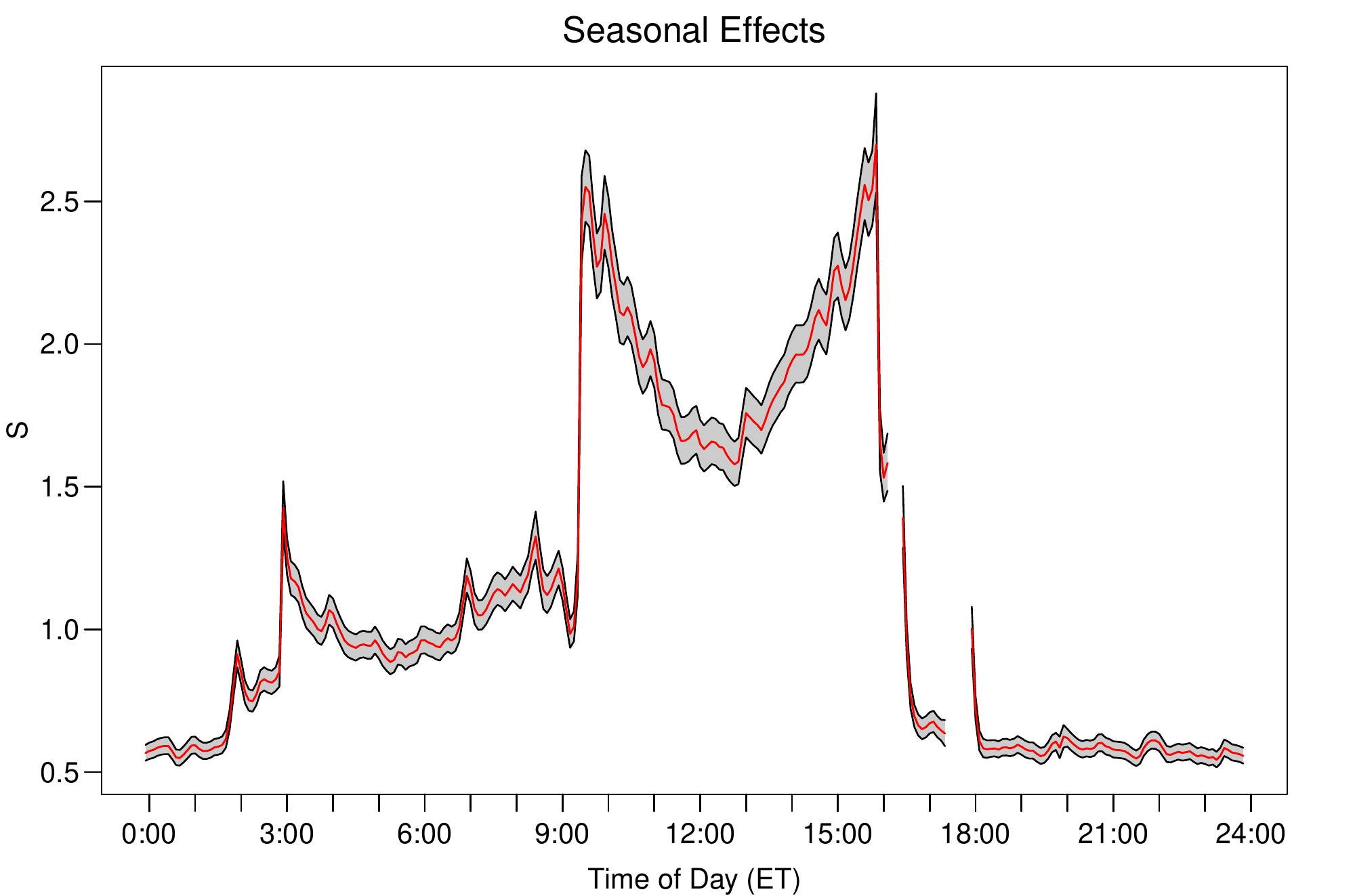} 
\end{center}
\caption{Posterior means and 95\% intervals for the seasonal effects, $%
\protect\beta=(\protect\beta_1,\ldots,\protect\beta_{288})$. Results are
shown on the standard deviation scale, $S=\exp(\protect\beta/2)$.
For example, a value of $S=2$ means that volatility is twice its baseline level.}
\label{fig:seasonal}
\end{figure}

Our jump estimates are `big,' as price jump volatility is about 4-8 times
unconditional 5-minute return volatilities. However, the sizes are
relatively small when compared to estimates from older daily price data or
option prices, which find rare jumps that are large and negative. Although
our sample contains some of the largest index moves ever observed in the
U.S. history, these were not large discontinuous moves, but rather a large
number of modest moves in the same direction. Thus, high-frequency data in
the most recent crisis provides a different view of jumps.

\begin{figure}[t]
\par
\begin{center}
\includegraphics[height=4.4in, width=6.35in]{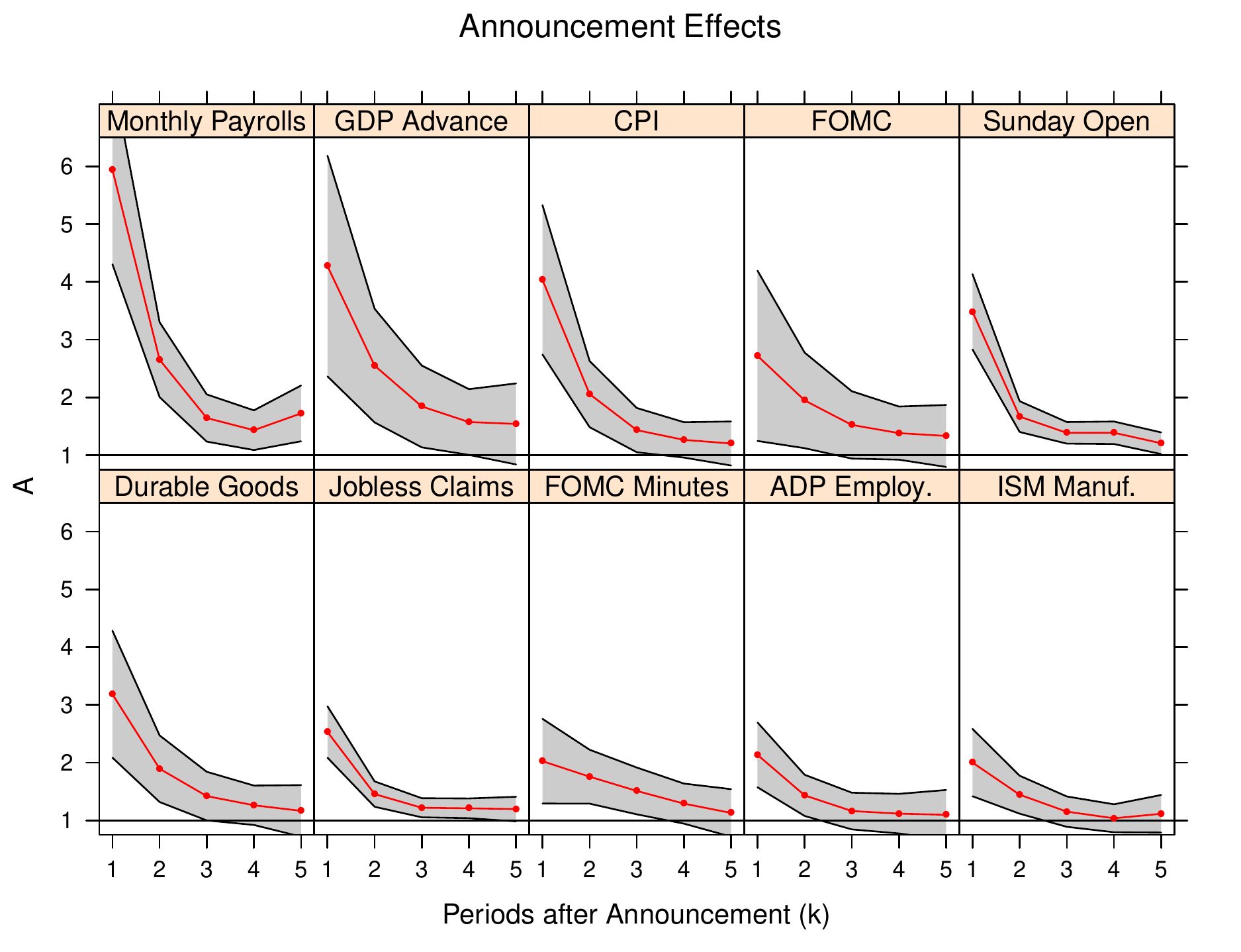}
\end{center}
\caption{Posterior means and 95\% intervals for the announcement effects, $%
\protect\alpha_i=(\protect\alpha_{i1}, \ldots,\protect\alpha_{i5})$. Results
are shown on the standard deviation scale, $A=\exp(\protect\alpha/2)$. 
For example, a value of $A=2$ means that volatility is twice its baseline level.}
\label{fig:announcements}
\end{figure}

Figure \ref{fig:seasonal} summarizes the posterior distribution of $S_{t}$. $%
S_{t}=1$ corresponds to average 5-minute volatility, so $S_{t}=0.5$ would
imply that volatility is roughly half average volatility. $S_{t}$ spikes to
more than 2.5 at the open and close of U.S. trading, and there is a clear
`U' shaped pattern during U.S. trading hours. $S_{t}$ fluctuates by a factor
of more than 5, highlighting the importance of predictable intraday
volatility.  Figure \ref{fig:announcements} summarizes the most important
announcements for the SVCJ$_{2}$ model (the other models are similar).
Volatility after Payrolls increases by 6 times, with the GDP, CPI and
FOMC announcements the next most important, with volatility increases of 
3-4 times.  The rate of decrease for the FOMC announcements are slower 
than for Payrolls, consistent with a greater digestion time.

\begin{figure}[tbp]
\par
\begin{center}
\includegraphics[height=7in, width=6in]{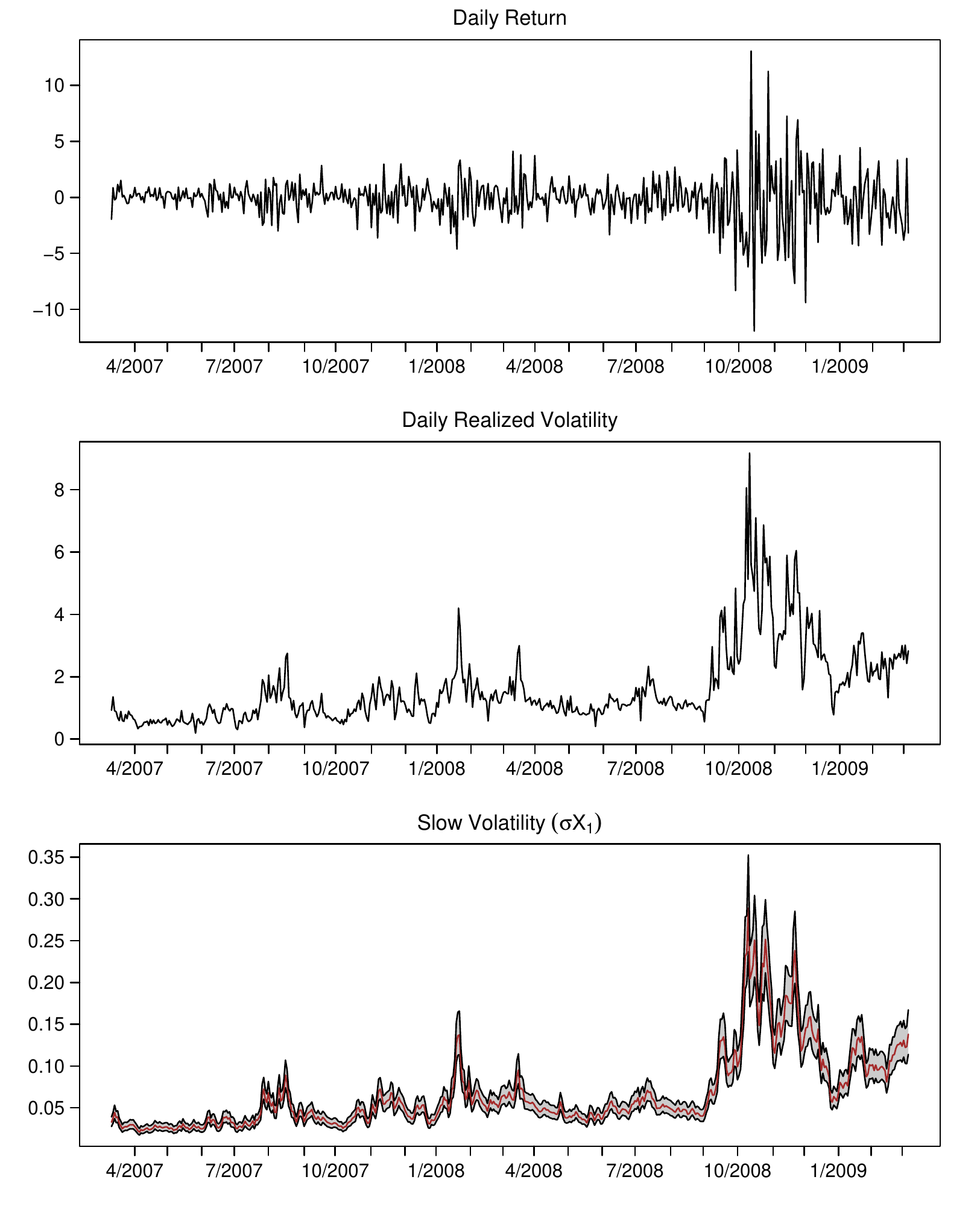}
\end{center}
\caption{Daily returns, realized volatility, and smoothed means and 95\%
intervals for the slow volatility component, $\protect\sigma X_1$, for the
SVCJ$_2$ model, March 2007--March 2009.}
\label{fig:fullsample}
\end{figure}

To understand interday volatility, Figure \ref{fig:fullsample} plots daily
returns, daily RV, and the slow volatility $\sigma X_{t,1}$. Volatility
spiked first in August 2007, with the panic in short-term lending markets.
Additional spikes occurred after the FOMC announcement in January 2008 and
the Bear Stearns takeover by J.P. Morgan in March 2008. Markets calmed down
until Fall 2008, when the crisis elevated volatility to its highest levels:\
on an annualized scale, $\sigma X_{t,1}$ was about 60\%. The slow factor
closely mirrors daily realized volatility.

\begin{figure}[tbp]
\begin{center}
\includegraphics[height=7.0535in,width=6.5in]{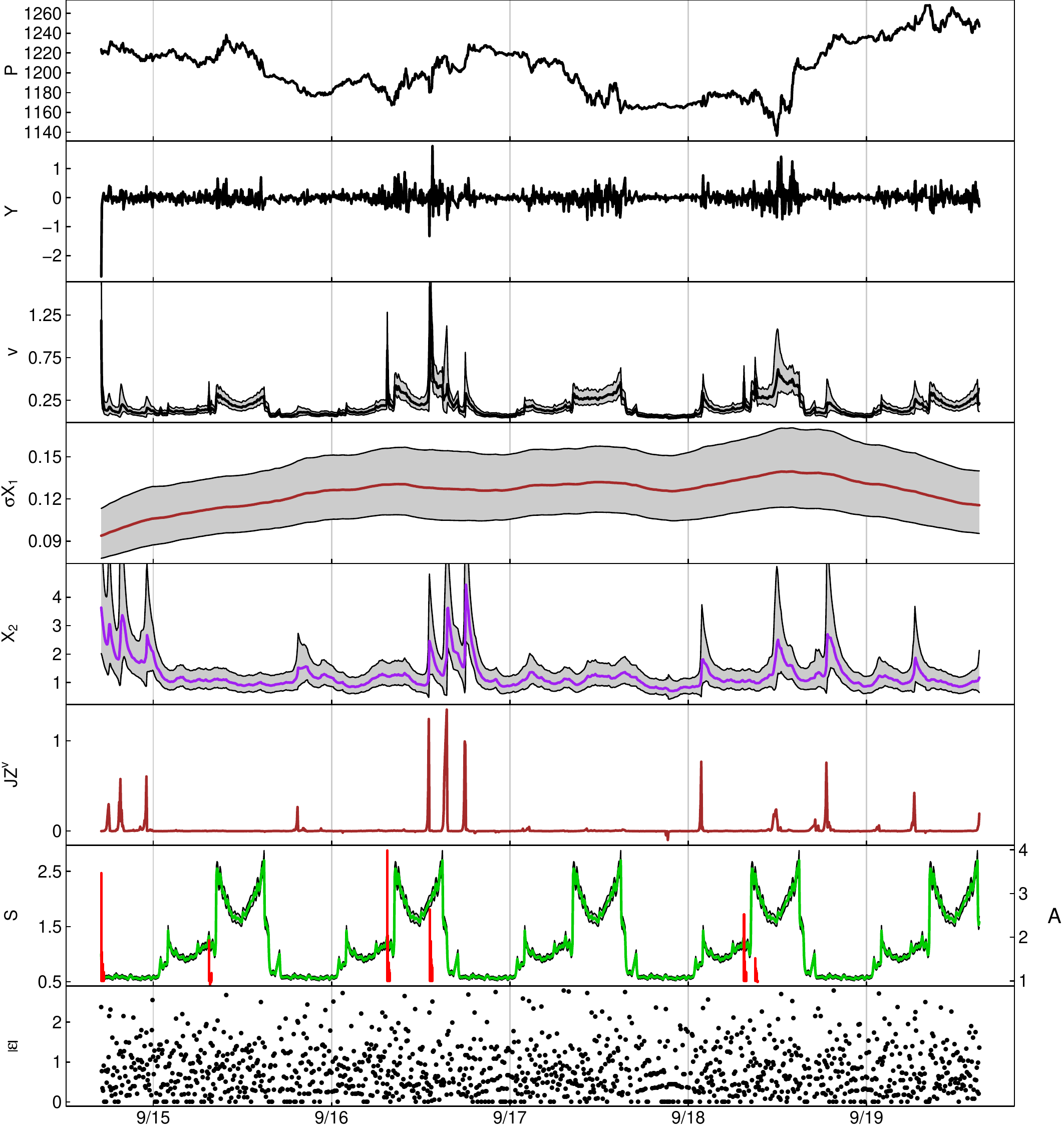}
\end{center}
\caption{Prices, returns, smoothed volatility components (total volatility,
slow volatility, fast volatility, volatility jumps, seasonal and
announcement components) and absolute value of the residuals during the week of September
14--19, 2008 for the SVCJ$_{2}$ model. Each panel contains posterior means,
and the bands represent 95\% posterior intervals. The second panel from the
bottom summarizes the seasonal fits on the left-hand axis and announcements
on the right. }
\label{fig:lehmansvcj2}
\end{figure}

To understand higher-frequency movements, Figure \ref{fig:lehmansvcj2} plots
the smoothed state variables during the week of September 14, 2008 for the SVCJ$_{2}$ model, when the
following happened:\ on September 14, Lehman Brothers filed for bankruptcy; on 
September 15, a large money market fund `broke the buck'; on September 16, AIG 
was bailed out, there was an FOMC meeting, and Bank of America announced their 
purchase of Merrill Lynch; and on September 18, the SEC banned short-selling of financial
stocks. The Sunday night overnight return was -2.75\%, as markets digested
the Lehman news. The model captures this move via a jump and elevated
intraday and interday volatility--interday volatility was more than twice
its long run average. On September 16, an FOMC\ announcement generated huge
volatility with three 5-minute returns greater than 1\%. Despite the elevated
announcement volatility, the model still needed a large jump in volatility.
After the close of normal trading, there were additional volatility jumps
corresponding to the Merrill Lynch merger. The large moves on September 18 were
associated with rumors and the subsequent announcement of the short-selling
ban on financial stocks, drove futures roughly 100 points higher overnight.

These results show the key role played by jumps in volatility and the fast
volatility factor, capturing the impact of unexpected news arrivals by
temporarily increasing volatility. In the SVt$_{2}$ model, large outlier
shocks generated by the t-distributed errors play a similar role in
explaining these large moves. Diffusive volatility is not able to increase
rapidly enough to capture extremely large movements.

\section{Out-of-sample results and applications}

Although in-sample fits are important, the ultimate test is predictive and
practical:\ how well does the model fit future data and can the model be
used for practical applications? In terms of overall predictive ability,
Figure \ref{fig:BIC1}b reports out-of-sample likelihood ratios relative to
the SV$_{1}$ model, which are based on the entire predictive distribution
and provide an overall measure of model fit. The ranking is nearly identical
to the in-sample results, and the GARCH models perform very poorly
out-of-sample in fitting the entire return distribution. This is strong
confirmation of model performance. In terms of applications, we consider
three (volatility forecasting, quantitative risk management, and a simple
volatility trading example) that are described below.

\subsection{Volatility forecasts}

Volatility forecasting is required for nearly every financial application, as
mentioned earlier, and is the gold-standard for evaluating estimators and
models when using intraday data \citep[see][]{AndeBenz:09}.
We compare volatility forecasts from our SV models to a range of GARCH and
nonparametric RV based estimators. We estimate parameters as of March 2009 and
forecast volatility from March 2009 to March 2012, a challenging period for three
reasons:\ the in-sample period is shorter than the out-of-sample period; the
out-of-sample period had lower volatility; and we do not update parameters
estimates.

We compute model based estimates, $\widehat{RV}_{s,\tau}^2,$ of realized
variance, $RV_{s,\tau }^{2}=\sum\nolimits_{t=1}^{\tau }y_{s+t}^{2},$ at
hourly ($\tau =12$) and daily ($\tau =279)$ horizons. The 5-minute forecasts
are similar to the hourly ones and are not reported. Table \ref{RVdaily}
reports forecast bias, mean-absolute forecasting errors (MAE), and 
forecasting regression $R^{2}$'s from Mincer-Zarnowitz regressions,%
\begin{equation*}
RV_{s,\tau }=b_{0}+b_{1}\,\widehat{RV}_{s,\tau }+\varepsilon _{s,\tau }.
\end{equation*}

\input test-rv

The SV models outperform all competitors. Compared to intraday GARCH, the
SV\ models provide a lower bias, lower MAE, and higher R$^{2}$'s. The SV
models generate daily R$^{2}$'s of 73\%, an almost 50\% improvement compared
to R$^{2}$'s of 47\% to 57\% for the GARCH specifications. This is a
remarkably high level of predictability. At hourly horizons, R$^{2}$, are
more than 10\% higher (e.g., R$^{2}$'s from 56\%-60\% to 66\%). All of the
SV models provide broadly similar fits, indicating that differences in
log-likelihoods are largely due to tail fits. We also benchmark to the
RV-based long-memory autoregressive (AR-RV) model of \cite{AndeBollDieb:03}, 
and the Realized GARCH model of \cite{HansHuanShek:12}.
These competitors are computed only at the daily horizon, following the
literature. Our SV models generate higher R$^{2}$'s in every case, and the
SV models' MAE and bias are generally similar or lower. The RV based models
clearly outperform the basic GARCH models.

\input test-reg

To attach statistical significance, we run bivariate `horse-race'
regressions,
 \begin{equation}
RV_{s,\tau }=b_{0}+b_{1}\,\widehat{RV}_{s,\tau }+b_{2}\,\widehat{RV}_{s,\tau
}^{SVCJ}+\varepsilon _{s,\tau },
\label{eqn:horse-race}
\end{equation}%
where $\widehat{RV}_{s,\tau }$ is from a competitor model and $\widehat{RV}%
_{s,\tau }^{SVCJ}$ is from the SVCJ$_2$ model. Table 6 summarizes the results.
Hourly, SVCJ$_2$ forecasts are highly significant (t-statistics greater than 50) 
in every case, and the competitors are insignificant in every case. The SVCJ$%
_{2}$ coefficients are close to but slightly less than one, and GARCH
coefficients are near zero. Daily SVCJ$_{2}$ forecasts are also highly
significant in every case, with t-statistics ranging from 12 to almost 30.
Interestingly, competitor forecasts are significant in many cases, though
less so than the SVCJ$_{2}$ forecasts. Economically, $b_{2}$ estimates are
close to one and those for the competitor models are close to zero. There is
some incremental information in some of the other models, as they are
significant in a number of cases, which suggests there is additional
predictability to be harvested. It would be interesting to consider an SV
model that treats lagged RV as a `regressor' variable, in a manner similar
to the Realized GARCH model.

Overall, the results provide additional confirmation to \cite{HansLund:05}'s 
important paper, which finds that it is possible to 
outperform simple GARCH(1,1) models. Parametric SV models provide strong 
improvements in forecasting ability, even in challenging periods of time.

\input test-ret1

\subsection{Risk management}

Quantitative risk management requires models to accurately fit
distributional tails in order to assess the risks of extreme losses.
Regulators often mandate VaR-based risk management procedures, which are
essentially real-time tail forecasts \citep[see, e.g.,][]{DuffPan:97}.
VaR is the loss in value that is exceeded with probability $p$, essentially 
the `$100-p^{th}$\%' critical value of the predictive distribution of returns.
Financial institutions compute VaR at daily or lower frequencies, but
intraday measures are useful for market makers, high frequency trading, and
options traders. To gain intuition, Figure 7 plots realized daily returns and 
the 1\% and 5\% daily VaR for the SVCJ$_{2}$ model. VaR
ranges from a low of well less than 1\% to a high of almost 20\% during the
crisis, with few noticeable or dramatic violations.

\begin{figure}[t]
\par
\begin{center}
\includegraphics[width=5.5in]{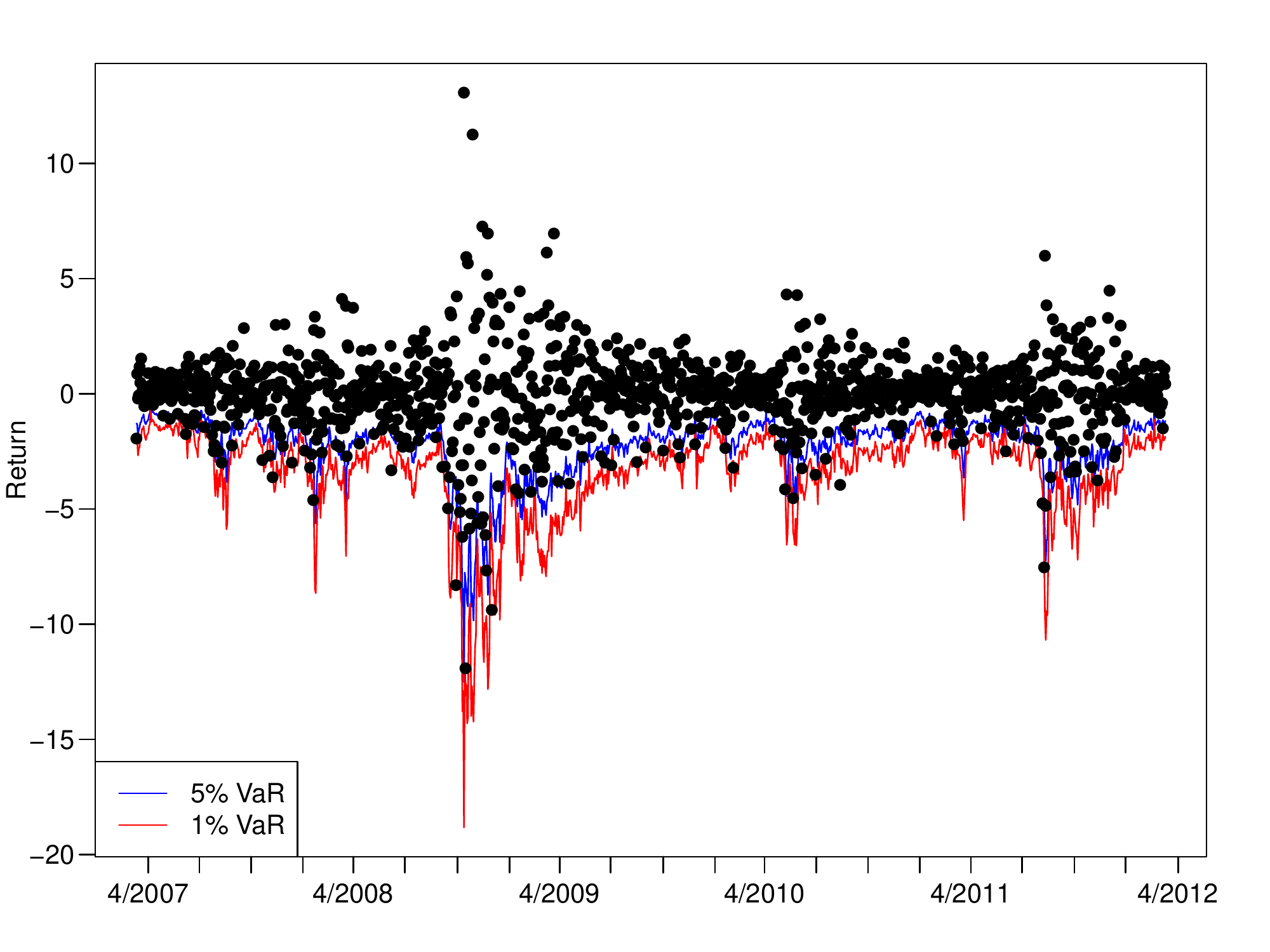}
\end{center}
\caption{Daily returns and out-of-sample 1\% and 5\% Value-at-Risk (VaR)
for daily returns for the SVCJ$_2$ model.}
\end{figure}

To evaluate the VaR performance out-of-sample, Table 7 reports 5-minute,
1-hour, and daily tail coverage probabilities at the 1\%, 5\%, and 10\%
levels, as well as a measure of total fit, $D$, which compares the ordered
predictive quantiles of the model with observed data. The SV models generate
more stable (across critical values and time horizons) and generally more
accurate VaR forecasts and distributional fits, with the SVCJ$_{2}$ model
performing marginally the best. Occasionally, a competitor model may perform
better at one frequency and for some quantiles, but no model uniformly
dominates the SV models. For example, the EGARCH-t model has the best
5-minute VaR performance, but it provides the worst at the daily frequency
and performs poorly in volatility forecasting. In terms of non-GARCH
competitors, the AR-RV models, due to a lack of return distribution, cannot
be used for VaR calculations. The RealGARCH models do not provide intraday
forecasts and, overall, the daily RealGARCH VaR statistics are generally on
par or slightly worse than the best performing SV\ models--slightly worse at
1\% level, better at the 5\% level, worse at the 10\% level, and worse in
terms of overall fit.

Overall, the multiscale SV models provide a robust and stable fit to the
tails of the return distribution over all horizons, which indicates their
potential usefulness for VaR based risk management.

\subsection{Volatility trading}

Volatility forecasts are useful for a range of practical applications, as
mentioned earlier. Documenting the economic benefits of a forecasting method
is, however, quite difficult, as most applications require additional
assumptions. For example, portfolio applications typically require expected
return estimates and a model of investor preferences, both of which are
arguably more difficult than volatility forecasting. This generates a
difficult joint specification problem:\ if, e.g., a trading strategy does
not work well, is it due to the volatility forecasts or the other components
of the problem?\ The same holds for derivatives pricing, as one must specify
risk premia and figure out how investors jointly learn about volatility from
derivatives pricing and historical returns. Because of this, few papers
analyze truly out-of-sample portfolio problems 
\cite*[see][for a review]{JohaKortPols:14}.

To highlight the economic value of our models while avoiding these
complexities, we implement a mean-reverting trading rule. Volatility is not
directly tradeable, and neither is the VIX\ index, but we base our trading
strategy on an ETF, the VXX, which is linked to futures on the VIX index.
Our trading strategy is based on volatility extremes and compares volatility
forecasts from various models with the VIX index, a common measure of option
implied volatility. For each model, we compute 5\% and 95\% predictive bands
for RV, either analytically (AR-RV and RealGARCH) or via simulation (for the
intraday models). If the VIX index is higher or lower than the 95\% or 5\%
bands, respectively, we enter into mean-reversion trades in the VXX, an ETF
inversely linked to the VIX index. If VIX crosses the median forecast (which
changes dramatically over time), we close the position. 
It is important to note that the procedure is fully out-of-sample and
applied symmetrically to all models.

\input vxx-trading-table

The financial crisis provides an interesting laboratory since it is likely
that any market inefficiencies or predictability might be magnified and thus
mean-reversion trades are a natural strategy to consider. This approach has
a number of other advantages:\ (1) it is a simple trading rule, (2)\ it
depends crucially on volatility forecasts, and (3) it allows for a direct 
\textit{relative} comparison of different forecasting models. Other
research, e.g., \cite{Nage:12} has documented the value of simple
mean-reversion trades during the crisis, suggestive of strong liquidity
premia or over-reaction.

Table 8 reports trade summaries for each model and a long/short portfolio that 
goes long the trades from the SVCJ$_{2}$ model and short those from other 
models.  Given the asymmetries in volatility--volatility tends to spike 
higher and mean-revert rapidly--there are many more long VXX trades (i.e., 
short the VIX index). The only exception is the EGARCH\ model, which at the 
daily level is strongly biased (see Table 5) and generates poor results. 
The other models generate positive annualized Sharpe ratios, indicative of 
predictive ability, but the multiscale SV models have higher Sharpe ratios 
than all of the competitor models. We also compute returns that are long the 
trading returns from the SVCJ$_{2}$ model and short another model. These 
Sharpe Ratios are always positive, and often on par with the Sharpe ratio for 
the SVCJ$_{2}$ model.  This essentially removes coincident trades and focuses 
on trades where the models disagree. This provides additional evidence for the 
practical utility of our approach.

\section{Conclusions}

This paper develops multifactor SV models of 24-hour intraday equity index
returns during and after the recent financial crisis. We estimate the models
directly using MCMC methods and use particle filtering methods for
forecasting and model evaluation. These models, more general than any in the
literature, provide a significant improvement in-sample and out-of-sample
fits, using both statistical metrics and applications.

In terms of model properties, we find strong evidence for multiscale
volatility, outliers (jumps or t-errors), periodic components capturing
intraday predictability, and announcements. Importantly, based on predictive
likelihoods, we find the exact same ordering of models in- and
out-of-sample, which indicates the results are robust and stable, even
during the extreme volatility realized in the crisis. Out-of-sample, we find
additional support for our approach based on superior volatility forecasts,
VaR risk management, which captures tail prediction, and a volatility
trading strategy.

These results are important as they document the practical usefulness of
sophisticated SV models for modeling intraday returns. Our results quantify
the improvements from carefully building models of intraday volatility that
account for jumps, multiple volatility factors, periodic components, and
announcements. SV models are not only competitive with RV or GARCH models,
but actually provide significant improvements in terms of in and
out-of-sample likelihood fits and applications like volatility forecasting,
risk management, and volatility trading.

\newpage

\bibliographystyle{rss}
\bibliography{acompat,references1}

\newpage

\section*{Appendix A: Model and Priors}

The general two-factor stochastic volatility model can be written as 
\begin{eqnarray*}
\mbox{Log Returns:} &&y_{t}=\mu +\exp (h_{t}/2)\sqrt{\lambda _{t}}%
\;\varepsilon _{t}+J_{t}Z_{t}^{y} \\
\mbox{Total Volatility:} &&h_{t}=\mu _{h}+x_{t,1}+x_{t,2}+f_{t}^{\prime
}\beta +I_{t}^{\prime }\alpha \\
\mbox{Slow Volatility:} &&x_{t,1}=\phi _{1}x_{t-1,1}+\sigma _{1}u_{t-1,1} \\
\mbox{Fast Volatility:} &&x_{t,2}=\phi _{2}x_{t-1,2}+\sigma _{2}\left( \rho
\varepsilon _{t-1}+\sqrt{1-\rho ^{2}}u_{t-1,2}\right) +J_{t-1}Z_{t-1}^{v} \\
\mbox{Periodic/Seasonal:} &&\beta _{i}\sim \mathcal{N}(0,\tau
_{s}^{2}U_{s})\cdot \mathbbm{1}(1^{\prime }\beta =0) \\
\mbox{Announcements:} &&\alpha _{i}\sim \mathcal{N}(0,\tau _{a}^{2}U_{a}) \\
\mbox{Scale Factors:} &&\lambda _{t}\sim \mathcal{IG}(\nu /2,\nu /2) \\
\mbox{Jump Times:} &&J_{t}\sim \mathcal{B}ern(\kappa ) \\
\mbox{Return Jumps:} &&Z_{t}^{y}\sim \mathcal{N}(\mu _{y},\sigma _{y}^{2}) \\
\mbox{Volatility Jumps:} &&Z_{t}^{v}\sim \mathcal{N}(\mu _{v},\sigma
_{v}^{2})
\end{eqnarray*}%
$U_{s}$ and $U_{a}$ are correlation matrices corresponding to the
cubic smoothing spline priors, as defined in Appendix B. We assume the
following prior distributions for the parameters: $\mu \sim \mathcal{N}(0,1)$%
, $\mu _{h}\sim \mathcal{N}(-6.2,1)$, $(\phi _{i}+1)/2\sim \mathcal{B}%
(20,1.5)$ (with $\phi _{1}>\phi _{2}$), $\sigma _{i}^{2}\sim \mathcal{IG}%
(.001,.001)$ for $i=1,2$, $\nu \sim \mathcal{DU}(2,128)$, $\rho \sim 
\mathcal{U}(-1,1)$, $\kappa \sim \mathcal{B}(1,1000)$, $(\mu _{y},\sigma
_{y}^{2})\sim \mathcal{NIG}(0,1,10,.324)$, $(\mu _{v},\sigma _{v}^{2})\sim 
\mathcal{NIG}(.50,10,10,1),$ and $\tau _{i}^{2}\sim \mathcal{IG}(.001,.001)$%
, for $i=s,a$, where $\mathcal{DU}$ is the discrete uniform, and $\mathcal{%
NIG}$ is the normal-inverse gamma distribution. The prior distribution for
$\phi _{i}$ is the transformed Beta distribution
proposed by \cite{ChibNardShep:02}.

\subsection*{Appendix B: Auxiliary Mixture Model}

We update the SV states and parameters using the mixture approximation of
\cite*{OmorChibShep:07}.  Conditional on $\mu ,J_{t},Z_{t}^{y},\lambda _{t}$, we
transform the returns to $(y_{t}^{\ast },d_{t})$, where 
\begin{equation*}
y_{t}^{\ast }=\log \left( \frac{y_{t}-\mu -J_{t}Z_{t}^{y}}{\sqrt{\lambda _{t}%
}}+const\right) ^{2},~~d_{t}=\mbox{sign}\left( y_{t}-\mu
-J_{t}Z_{t}^{y}\right) ,
\end{equation*}%
and $const=.0001$ is used to avoid logs of zeros. We then write the return
equation as 
\begin{equation*}
y_{t}^{\ast }=h_{t}+\log (\varepsilon _{t}^{2}),
\end{equation*}%
and approximate the joint distribution of $\zeta _{t}=\log (\varepsilon
_{t}^{2})$ and $\eta _{t,2}$ by a mixture of 10 normals: 
\begin{equation*}
p(\zeta _{t},\eta _{t,2}|d_{t},\rho ,\sigma )=\sum_{j=1}^{10}\,p_{j}\,%
\mathcal{N}(\zeta _{t}|m_{j},v_{j}^{2})\,\mathcal{N}(\eta _{t,2}|d_{t}\rho
(a_{j}^{\ast }+b_{j}^{\ast }\zeta _{t}),1-\rho ^{2}),
\end{equation*}%
where $(p_{j},m_{j},v_{j},a_{j}^{\ast },b_{j}^{\ast }),j=1,\ldots ,10$ are
constants specified in \cite{OmorChibShep:07}. We then introduce a set of mixture
indicator variables $\omega _{t}\in \{1,\ldots ,10\}$ for $t=1,\ldots ,T$.
Conditional on the indicators, the model has a linear Gaussian state-space
form, and the FFBS algorithm is used to generate the volatility states 
and parameters.

\subsection*{Appendix C: Cubic Smoothing Splines}

To estimate the seasonal and announcement effects, we use the state-space
framework for polynomial smoothing splines of \cite{KohnAnsl:87}. Let $%
g=(g_1,\ldots,g_K)$ denote the unknown coefficients, which have a modified
cubic smoothing spline prior of the form $\nabla^2 g_k \sim \mathcal{N}(0,
c_k^2 \tau^2)$, where $c_k$ are known constants and $\tau^2$ is an unknown
smoothing parameter. We observe data $y_k \sim \mathcal{N}(g_k,v^2_k)$ for $%
k=1,2,\ldots,K$, where $v_k^2$ are known. If we define the state vector as $%
x_k=(g_k,\dot{g}_k)^{\prime }$, the model can be written in state-space form
as 
\begin{eqnarray*}
y_k &=& h^{\prime }x_k + \epsilon_k, ~~ \epsilon_k \sim \mathcal{N}(0,v^2_k)
\\
x_{k+1} &=& Fx_k + u_k, ~~ u_k \sim \mathcal{N}(0,\tau^2c_k^2U),
\end{eqnarray*}
where $h=(1,0)^{\prime }$, 
\begin{equation*}
F=\left(%
\begin{tabular}{cc}
1 & 1 \\ 
0 & 1%
\end{tabular}%
\right), ~~~ U=\left(%
\begin{tabular}{cc}
1/3 & 1/2 \\ 
1/2 & 1%
\end{tabular}%
\right),
\end{equation*}
$\epsilon_k$ and $u_k$ are serially and mutually uncorrelated errors, and $%
x_1 \sim \mathcal{N}(0,c_1^2I) $ with $c_1$ large. Defining $%
x=(x_1,\dots,x_K)$ and $y=(y_1,\ldots,y_K)$, and assuming the prior $\tau^2
\sim p(\tau^2)$, the posterior distribution of interest is 
\begin{equation*}
p(x,\tau^2|y) \propto p(\tau^2) \prod_{k=1}^K p(y_k|x_k)
p(x_{k+1}|x_k,\tau^2).
\end{equation*}
We use a Metropolis step to generate $x$ and $\tau^2$ jointly from this
distribution. Conditional on the current value, $\tau^{2(i)}$, draw $%
\tau^{2(*)} \sim \mathcal{N}(\tau^{2(i)},w)$, and accept with probability 
\begin{equation*}
\min\left\{1,\frac{p(y|\tau^{2(*)})\,p(\tau^{2(*)})}{p(y|\tau^{2(i)})\,p(%
\tau^{2(i)})}\right\}.
\end{equation*}
Here $p(y|\tau^2)$ is computed using the Kalman filter. If the draw is
accepted, set $\tau^{2(i+1)}=\tau^{2(*)}$ and generate $x^{(i+1)}\sim
p(x|\tau^{2(i+1)},y)$ using the FFBS algorithm. Otherwise leave $x$
unchanged. Since $x_k=(g_k,\dot{g}_k)$, draws of the function $%
g=(g_1,\ldots,g_K)^{\prime }$ are obtained directly from $x$.

The degrees of freedom for the fit is obtained by noting that the posterior
mean of the function, conditional on $\tau^2$, has the form $%
E(g|y,\tau^2)=Ay $, where $A$ is the so-called ``hat-matrix.'' The degrees
of freedom is defined as $d=tr(A)$. Following \cite{AnslKohn:87}, this
value is computed efficiently using a modified Kalman filter algorithm.

\section*{Appendix D: MCMC Algorithm}

The joint posterior distribution for the model in Appendix A is 
\begin{equation*}
p(x,\lambda,J,Z,\beta,\alpha,\theta|y) \propto
p(y|x,\lambda,J,Z,\beta,\alpha,\theta) p(x,\lambda,J,Z|\theta) \,
p(\beta|\theta) \, p(\alpha|\theta) \, p(\theta)
\end{equation*}
where $x_t=(x_{t,1},x_{t,2})$, $Z_t=(Z_t^y,Z_t^v)$, $y=(y_1,\ldots,y_T)$, $%
x=(x_1,\ldots,x_T)$, $\lambda=(\lambda_1,\ldots,\lambda_T)$, $%
J=(J_1,\ldots,J_T)$, $Z=(Z_1,\ldots,Z_T)$, and $\theta=(\mu,\mu_h,\phi_1,%
\phi_2,\sigma_1,\sigma_2,\rho,\nu,
\kappa,\mu_y,\sigma_y,\mu_v,\sigma_v,\tau_s,\tau_a)$.

\bigskip \noindent The models were estimated using the Markov chain Monte
Carlo algorithm described below. We ran the MCMC for 12,500 iterations and
discarded the first 2500 as burn-in, leaving 10,000 samples for posterior
inference.  For the SVCJ$_2$ model, we ran the chain for 1,000,000 
iterations after a burn-in of 25,000 iterations, and retained every 100th draw,
leaving 10,000 samples for inference.  Diagnostic plots and tests indicated 
no obvious problems with convergence.  The starting values were set to the 
prior mean or mode, although we found that the results were robust to this 
choice.  The MCMC algorithm followed by a description of the full conditional 
posterior distributions are given below.

\begin{my_enumerate}
\item Draw $p(\omega|y^*,x,J,Z,\lambda,\theta)$

\item Draw $p(x,\mu_h,\phi_i,\sigma_i,\rho|y^*,\omega,\beta,\alpha), i=1,2$

\item Draw $p(\beta,\tau_s^2|y^*,\omega,x,\alpha,\theta)$

\item Draw $p(\alpha,\tau_a^2|y^*,\omega,x,\beta,\theta)$

\item Draw $p(\lambda,\nu|y,x,J,Z,\beta,\alpha,\mu)$

\item Draw $p(J,Z|y,x,\lambda,\kappa,\mu_j,\sigma_j,\mu), j=y,v$

\item Draw $p(\kappa,\mu_j,\sigma_j | J,Z), j=y,v$

\item Draw $p(\mu|y,x,\lambda,J,Z,\beta,\alpha)$
\end{my_enumerate}

\begin{enumerate}
\item {\underline {\textbf{Sampling $\omega_t$}}.} The indicators $\omega_t$
are independent multinomials with probabilities 
\begin{equation*}
Pr(\omega_t=j|\zeta_t,\eta_{t,2},\rho) \, \propto \, p_j \, \phi(\zeta_t;
m_j,v_j^2) \, \phi(\eta_{t,2}; d_t\rho (a_j^*+b_j^*\zeta_t),1-\rho^2):
j=1,\ldots,10.
\end{equation*}

\item {\underline {\textbf{Sampling $(\mu, \phi_1,\phi_2, \sigma_1,
\sigma_2, \rho, x_{t,1}, x_{t,2})$}}.} Conditional on the other states and parameters
and the mixture indicators, the model can be written in state-space form: 
\begin{eqnarray*}
\hat{y}_t &=& \mu + x_{t,1} + x_{t,2} + v_{\omega_t} u_{t,1} \\
x_{t+1,1} &=& \phi_1 x_{t,1} + \sigma_1 u_{t,2} \\
x_{t+1,2} &=& \phi_2 x_{t,2} + \sigma_2 (d_t \rho
(a_{\omega_t}^*+b_{\omega_t}^*u_{t,1}) + \sqrt{1-\rho^2}u_{t,3}) + J_t Z_t^v
\end{eqnarray*}
where $\hat{y}_t=y_t^*-s_t-a_t-m_{\omega_t}$, and $%
(u_{t,1},u_{t,2},u_{t,3})^{\prime }\sim \mathcal{N}_3(0,I)$. We use 
\cite{OmorChibShep:07}'s method to draw the states and parameters 
$(\phi_1,\sigma_1,\phi_2,\sigma_2,\rho, \mu_h,x_{t,1},x_{t,2})$ as a block 
from the full conditional. To update the parameters $(\phi_1,\phi_2,\sigma_1,\sigma_2,\rho)$ 
we use a Metropolis step, with a truncated multivariate normal proposal distribution 
with covariance matrix chosen to achieve an acceptance probability of around 30\%.


\item {\underline {\textbf{Sampling $(\beta, \tau_s)$}}.} Conditional on the
other states and parameters, we cast the model in state-space form by
defining the state vector as $\beta_k^*=(\beta_k,\dot{\beta}_k)^{\prime }$,
and writing 
\begin{eqnarray*}
\hat{y}_k &=& h^{\prime }\beta_k^* + \epsilon_k, ~~~ \epsilon_k \sim 
\mathcal{N}(0,\hat{v}_k^2) \\
\beta_{k+1}^* &=& F \beta_k^* + u_k, ~~ u_k \sim \mathcal{N}(0,c_k^2\tau_s^2
U); ~~k=1,\ldots,288,
\end{eqnarray*}
where $\hat{y}_k= \hat{v}_k^2\sum_{\{t:f_{tk}=1\}}(y_t^*-%
\mu-x_{t,1}-x_{t,2}-a_t-m_{\omega_t})v_{\omega_t}^{-2}$, $\hat{v}_k^2
=(\sum_{\{t:f_{tk}=1\}}v_{\omega_t}^{-2})^{-1}$, and 
\begin{equation*}
c_k = \left\{ 
\begin{tabular}{ll}
$100$ & if $k = 1, 25, 109, 187, 265, 271$; \\ 
$1$ & otherwise,%
\end{tabular}
\right.
\end{equation*}
are variance inflation factors used to generate discontinuities at
market opening/closing times. We then use the Metropolis algorithm from
Appendix C to generate $(\tilde\beta,\tau_s^2)$ as a block from the unconstrained
posterior. We impose the zero-sum constraint on the seasonal coefficients by 
setting $\beta_k = \tilde{\beta}_k - (\sum_{k=1}^{288} \tilde{\beta}_k)/288$.

\item {\underline {\textbf{Sampling $(\alpha, \tau_a)$}}.} For each
announcement type $i=1,\ldots,n$, the model is cast in state-space form by
defining the state vector as $\alpha_{i,k}^*=(\alpha_{i,k},\dot{\alpha}%
_{i,k})^{\prime }$ and writing 
\begin{eqnarray*}
\hat{y}_{ik} &=& h^{\prime }\alpha_{ik}^* + \epsilon_{ik}, ~~~ \epsilon_{ik}
\sim \mathcal{N}(0,\hat{v}_{ik}^2) \\
\alpha_{i,k+1}^* &=& F \alpha_{ik}^* + u_{ik}, ~~ u_{ik} \sim \mathcal{N}%
(0,\tau_s^2U); ~~k=1,\ldots,5,
\end{eqnarray*}
where $\hat{y}_{ik} = \hat{v}_{ik}^2
\sum_{\{t:I_{tik}=1\}}(y_t^*-\mu-x_{t,1}-x_{t,2}-s_t-m_{\omega_t})v_{%
\omega_t}^{-2}$ and $\hat{v}_{ik}^2 =
(\sum_{\{t:I_{tik}=1\}}v_{\omega_t}^{-2})^{-1}$. We then use the Metropolis
algorithm from Appendix B to update $(\alpha,\tau_a^2)$.

\item {\underline {\textbf{Sampling $(\lambda,\nu)$}}.} Write the joint
posterior as $p(\lambda_,\nu|\mbox{rest}) = p(\nu|\mbox{rest}) p(\lambda_t|\nu,\mbox{rest})$. To
update the degrees of freedom, define $w_t = (y_t-\mu-J_tZ_t^y)/V_t$, so 
the model is $(w_t|\nu,\mbox{rest}) \sim t_\nu(0,1)$. Under the discrete
uniform prior $\nu \sim \mathcal{DU}(2,128)$, the posterior is a multinomial
distribution $(\nu|w,\mbox{rest}) \sim \mathcal{M}(\pi_2^*,\ldots,\pi_{128}^*)$,
with probabilities 
\begin{equation*}
\pi_{\nu}^* \propto \prod_{t=1}^T p_{\nu}(w_t), ~~~ \nu=2,\ldots,128,
\end{equation*}
where $p_\nu(\cdot)$ denotes the Student-$t$ density with $\nu$ degrees of
freedom. Rather than compute each of the multinomial probabilities, which is
very costly, we use a Metropolis step to update $\nu$. Given the current
value $\nu^{(i)}$, we draw a candidate value $\nu^{(*)} \sim \mathcal{DU}%
(\nu^{(i)}-\delta,\nu^{(i)}+\delta)$, and accept with probability 
\begin{equation*}
\min\left\{1,\frac{\prod_{t=1}^T p_{\nu^{(*)}}(w_t)}{\prod_{t=1}^T
p_{\nu^{(i)}}(w_t)}\right\}.
\end{equation*}
The width $\delta$ is chosen to give an acceptance probability between 20\%
and 50\%.

To update the scale factors, define $\varepsilon_t^* = (y_t - \mu -
J_tZ_t^y)/\sqrt{V_t}$. Then $(\varepsilon_t^*|\lambda_t,\nu,\mbox{rest}) \sim 
\mathcal{N}(0,\lambda_t)$. Combining this with the prior, $(\lambda_t|\nu)
\sim \mathcal{IG}(\nu/2,\nu/2)$, the full conditional is 
\begin{eqnarray*}
(\lambda_t \,|\, \nu,\mbox{rest}) & \sim & \mathcal{IG}\left(\frac{\nu+1}{2},\frac{%
\nu+\varepsilon_t^{*2}}{2}\right).
\end{eqnarray*}

\item {\underline {\textbf{Sampling $(J, Z^y, Z^v)$}}.} Given the other
states and parameters, write the model as $(w_t|J_t,Z_t,\mbox{rest}) \sim \mathcal{N}%
(J_tZ_t,\Sigma_t)$, where 
\begin{equation*}
w_t = \left(%
\begin{tabular}{c}
$y_t-\mu$ \\ 
$x_{t+1,2}-\phi_2x_{t,2}$%
\end{tabular}%
\right), ~~ Z_t = \left(%
\begin{tabular}{c}
$Z_t^y$ \\ 
$Z_t^v$%
\end{tabular}%
\right), ~~ \Sigma_t = \left(%
\begin{tabular}{cc}
$\lambda_tV_t$ & $\rho\sigma_2\sqrt{\lambda_tV_t}$ \\ 
$\rho\sigma_2\sqrt{\lambda_tV_t}$ & $(1-\rho^2)\sigma_2^2$%
\end{tabular}%
\right),
\end{equation*}
Assuming conjugate priors, $J_t \sim \mathcal{B}ern(\kappa)$ and $Z_t\sim 
\mathcal{N}(\mu_z,\Sigma_z)$, where $\mu_z=(\mu_y,\mu_v)^{\prime }$ and $%
\Sigma_z=\mbox{diag}(\sigma^2_y,\sigma^2_v)$, the full conditionals for the
jump times and sizes are 
\begin{equation*}
P(J_t=1 \,|\, \mbox{rest}) = \frac{\kappa \,
\phi\left(w_t;\mu_z,\Sigma_t+\Sigma_z\right)} {(1-\kappa) \,
\phi\left(w_t;0,\Sigma_t\right) + \kappa \,
\phi\left(w_t;\mu_z,\Sigma_t+\Sigma_z\right)}
\end{equation*}
\begin{equation*}
(Z_t | J_t =1, \mbox{rest}) \sim \mathcal{N}\left((\Sigma_z^{-1} +
\Sigma_t^{-1})^{-1}(\Sigma_z^{-1}\mu_z + \Sigma_t^{-1}w_t),(\Sigma_z^{-1} +
\Sigma_t^{-1})^{-1}\right).
\end{equation*}

\item {\underline {\textbf{Sampling $(\kappa,\mu_y,\sigma_y, \mu_v,\sigma_v)$%
}}.} Under the conjugate priors $\kappa \sim \mathcal{B}(a_\kappa,b_\kappa)$%
, and $(\mu_j,\sigma_j^2) \sim \mathcal{NIG}(m_j,c_j,a_j,b_j)$, for $j=y,v$,
the full conditionals for the jump parameters are
\begin{eqnarray*}
(\kappa \,|\, \mbox{rest}) & \sim & \mathcal{B}(a_\kappa^*,b_\kappa^*) \\
(\mu_{y},\sigma_{y}^2 \,|\, \mbox{rest}) & \sim & \mathcal{NIG}(m_y^*,c_y^*,a_y^*,b_y^*) \\
(\mu_v,\sigma^2_v \,|\, \mbox{rest}) & \sim & \mathcal{NIG}(m_v^*,c_v^*,a_v^*,b_v^*)
\end{eqnarray*}

\vspace{-.5cm} where for $j=y,v$, 
\begin{eqnarray*}
a_\kappa^* &=& a_\kappa + \sum\nolimits_{t=1}^T J_t,~~~~~ b_\kappa^* =
b_\kappa + T - \sum\nolimits_{t=1}^T J_t \\
c_j^* &=& c_j + \sum\nolimits_{t=1}^T J_t, ~~~~~ c_j^*m_j^* = c_j
m_j+\sum\nolimits_{t=1}^T J_tZ_t^j \\
a_j^* &=& a_j + \sum\nolimits_{t=1}^T J_t, ~~~~~ b_j^* = b_j+c_j
m_j^2+\sum\nolimits_{t=1}^T (J_tZ_t^j)^2-c_j^*m_j^{*2}.
\end{eqnarray*}

\item {\underline {\textbf{Sampling $\mu$}}.} Under the prior distribution, $%
\mu\sim \mathcal{N}(m_\mu,v_\mu)$, the full conditional for the mean return
is 
\begin{equation*}
(\mu \,|\, \mbox{rest}) \sim \mathcal{N} \left( \left(\frac{1}{v_\mu} + \sum_{t=1}^T 
\frac{1}{V_t^*}\right)^{-1} \left(\frac{m_\mu}{v_\mu} + \sum_{t=1}^T\frac{%
y_t^*}{V_t^*}\right), \left(\frac{1}{v_\mu} + \sum_{t=1}^T \frac{1}{V_t^*}%
\right)^{-1} \right)
\end{equation*}
where $V_t^*=(1-\rho^2)\lambda_tV_t$ and $y_t^* = y_t - J_tZ_t^y - \frac{\rho%
\sqrt{\lambda_t V_t}}{\sigma_2}\left(x_{t+1,2}-\phi_2x_{t,2}-J_tZ_t^v\right)$%
. 
\end{enumerate}


\section*{Appendix E: Auxiliary Particle Filter}

We describe a general auxiliary particle filter used for the models in the
paper. Assume the parameters are fixed at their posterior means. Write the
state vector as $z_{t}=(x_{t},z_{t}^{\ast })$, where $x_{t}$ are the SV
states and $z_{t}^{\ast }$ are the other states (e.g., jump times, jump sizes, etc.)
in the model. Our goal is to sequentially sample from the filtering distribution $p(z_{t}|y^{t})$, 
for $t=1,\ldots,T$.  In our models, it is convenient to factorize this 
distribution as $p(z_{t}|y^{t})=p(x_{t}|y^{t})p(z_{t}^{\ast }|x_{t},y^{t})$, 
where the first distribution on the right side is unavailable
analytically, and second is available in closed form.

Assume we have an equally-weighted sample available at time $t-1$, $%
z_{t-1}^{(i)} \sim p(z_{t-1}|y^{t-1})$, for $i=1,\ldots,N$. We use the
auxiliary particle filter to sample from the joint distribution $%
p(k,x_t|y^t) \propto p(k|y^{t-1})p(x_t|z_{t-1}^{(k)},y^{t-1}) p(y_t|x_t)$,
where $k$ is the auxiliary mixture index. To do this, we first draw the
index $k \sim q(k|y^t)$ and then the state, $x_t \sim q(x_t|k,y^t) $, where 
\begin{eqnarray*}
q(k|y^t) &\propto & p\left(y_t|\hat{x}_t^{(k)}\right), \\
q(x_t|k,y^t) &=& p\left(x_t|z_{t-1}^{(k)},y^{t-1}\right),
\end{eqnarray*}
and $\hat{x}_t^{(k)} = E(x_t|z_{t-1}^{(k)},y^{t-1})$. We then resample the
states with weights 
\begin{equation*}
w_t^{(i)} \propto \frac{p\left(y_t|x_t^{(i)}\right)} {p\left(y_t|\hat{x}%
_t^{(k^i)}\right)}.
\end{equation*}
to obtain samples from the posterior distribution, $p(x_t|y^t)$. We then
draw from $p(z_t^*|x_t,y^t)$ which is available analytically. 
This leads to the following APF algorithm. 

\begin{my_enumerate}
\item Start with a sample $z_{t-1}^{(i)}
=\left(x_{t-1}^{(i)},z_{t-1}^{*(i)}\right)\sim p(z_{t-1}|y^{t-1})$.

\item Compute $\pi_t^{(i)} \propto p\left(y_t|\widehat{x}_t^{(i)}\right)$,
where $\widehat{x}_t^{(i)} = E\left(x_t|z_{t-1}^{(i)},y^{t-1}\right)$.

\item Generate $k^i \sim \mathcal{M}\left(\pi_t^{(1)},\ldots,\pi_t^{(N)}%
\right)$.

\item Generate $\tilde{x}_t^{(i)} \sim
p\left(x_t|z_{t-1}^{(k^i)},y^{t-1}\right)$.

\item Compute $w_t^{(i)} \propto p\left(y_t|\tilde{x}_t^{(i)}\right)/%
\pi_t^{(k^i)}$.

\item Generate $j^i \sim \mathcal{M}\left(w_t^{(1)},\ldots,w_t^{(N)}\right)$
and set $x_t^{(i)}=\tilde{x}_t^{(j^i)}$

\item Generate $z_t^{*(i)} \sim p\left(z_t^*|x_t^{(i)},y^t\right)$.
\end{my_enumerate}

\vspace{.15cm} \noindent Following \cite{MaliPitt:12}, the likelihood
function for a fixed parameter value $\Theta$ can be approximated using the
output from the auxiliary particle filter as 
\begin{equation*}
\mathcal{L}(y^T|\Theta) = \prod_{t=1}^T \left(\frac{1}{N} \sum_{i=1}^N
\pi_t^{(i)}\right) \left(\frac{1}{N} \sum_{i=1}^N w_t^{(i)}\right).
\end{equation*}

\vspace{.5cm} \noindent As an illustration of the APF algorithm, consider the 
SVCJ$_2 $ model. For this model, we define $x_t=(x_{t,1},x_{t,2})$, 
$z_t^*=(J_t,Z_t^y,Z_t^v) $, $V_t = \exp(h_t)$, and $\varepsilon_t =
(y_t-\mu-J_tZ_t^y) /\sqrt{V_t}$, and the conditional distributions 
used for the algorithm are 
\begin{itemize}
\item $p(x_t|z_{t-1},y^{t-1}) = \mathcal{N}\left( \left(%
\begin{tabular}{c}
$\phi_1x_{t-1,1}$ \\ 
$\phi_2x_{t-1,2}+\sigma_2\rho\varepsilon_{t-1}+J_{t-1}Z_{t-1}^v$%
\end{tabular}
\right), \left(%
\begin{tabular}{cc}
$\sigma^2_1$ & 0 \\ 
0 & $\sigma^2_2(1-\rho^2)$%
\end{tabular}
\right) \right)$

\item $p(y_t|x_t) = (1-\kappa)\, \phi(y_t;\mu,V_t) + \kappa \,
\phi(y_t;\mu+\mu_y,V_t+\sigma^2_y)$ 

\item $p(J_t=1|x_t,y_t) = \kappa \,
\phi\left(y_t;\mu+\mu_y,V_t+\sigma^2_y\right)/p(y_t|x_t) 
$

\item $p(Z_t^y|J_t,x_t,y^t) = \mathcal{N}\left( \left(\frac{1}{\sigma_y^2}+%
\frac{J_t}{V_t}\right)^{-1} \left(\frac{\mu_y}{\sigma_y^2}+\frac{J_t(y_t-\mu)%
}{V_t}\right), \left(\frac{1}{\sigma_y^2}+\frac{J_t}{V_t}\right)^{-1}\right)$

\item $p(Z_t^v|J_t,x_t,y^t) = \mathcal{N}\left(\mu_v,\sigma^2_v\right)$ 
\end{itemize}

\section*{Appendix F: Forecasting Returns \& Realized Volatility}

Conditional on posterior samples of the states at time $s$, $z_s^{(i)} \sim
p(z_s|y^s), i=1,\ldots,N$, and fixed parameter values, we approximate the
forecast distribution of returns and realized volatility by forward
simulation. We forecast over a $\tau$-period horizon as follows. 
For $i=1,\ldots,N$ and $t=1,\ldots,\tau$ we generate 
$z_{s+t}^{(i)} \sim p(z_{s+t}|z_{s+t-1}^{(i)})$ and 
$y_{s+t}^{(i)} \sim p(y_{s+t}|z_{s+t}^{(i)})$.  We then aggregate the simulated 
returns and squared returns to obtain samples of returns and realized
volatility at horizon $\tau$:
\begin{equation*}
y_{s,\tau}^{(i)} = \sum_{t=1}^{\tau} y_{s+t}^{(i)} ~~~~ \mbox{and} ~~~~ 
\widehat{RV}_{s,\tau}^{(i)} = \sqrt{\sum_{t=1}^{\tau}
\left(y_{s+t}^{(i)}\right)^2}.
\end{equation*}
The empirical 1\%, 5\%, 10\% quantiles of the return distribution are used
to estimate Value-at-Risk.  The point prediction for RV is obtained as the
posterior mean (across simulations):
\begin{equation*}
\widehat{RV}_{s,\tau} = \frac{1}{N}\sum_{i=1}^N \widehat{RV}_{s,\tau}^{(i)}.
\end{equation*}

\newpage
\section*{Appendix G: GARCH Models}

\begin{itemize}

\item {\bf Intraday GARCH Models with Seasonality.}
Let $y_t$ denote the 5-minute return, $s_t$ denote the seasonal effect, and $\sigma_t$ 
the unobserved volatility at period $t$.  Our intraday GARCH models assume one of the 
following return equations (either normal or t):
\begin{eqnarray*}
\mbox{Normal}: && y_t = s_t \sigma_t z_t   ~~~ z_t \sim \mathcal{N}(0,1)\\
\mbox{Student-t}: && y_t = \sqrt{\frac{\nu-2}{\nu}} s_t \sigma_t z_t,  ~~~ z_t \sim t_\nu(0,1)
\end{eqnarray*}
Define the seasonally-adjusted returns as $\tilde{y}_t = y_ts_t^{-1}$ for the normal model,
and as $\tilde{y}_t =\sqrt{\nu/(\nu-2)}y_ts_t^{-1}$ for the student-t model.  
We can then write the model as $\tilde{y}_t=\sigma_tz_t$. 
We consider three different GARCH models for the dynamics of $\sigma_t$:
\begin{eqnarray*}
\mbox{GARCH}:  && \sigma_t^2 = \omega + \alpha \tilde{y}_{t-1}^2 + \beta \sigma_{t-1}^2 \\ 
\mbox{GJR}:    && \sigma_t^2 = \omega + \alpha \tilde{y}_{t-1}^2 + \gamma\tilde{y}_{t-1}^2I(\tilde{y}_{t-1}<0)+ \beta \sigma_{t-1}^2\\
\mbox{EGARCH}: && \log(\sigma_t^2) = \omega + \alpha (|z_{t-1}|-E(|z_{t-1}|)) + \gamma z_{t-1} + \beta \log (\sigma_{t-1}^2)
\end{eqnarray*}
The model is estimated in two stages. 
Define $s_t=f_t\beta$, where 
$f_t=(f_{t,1},\ldots,f_{t,288})'$, where $f_{tk}=1$ if time $t$ is 
period of the day $k$ and zero otherwise, and $\beta=(\beta_1,\ldots,\beta_{288})'$.
The seasonal coefficients are estimated by
\begin{equation*}
\hat{\beta}_k^2  = \frac{\sum_{t:f_t=k} y_t^2}{N_k}, ~~ k=1,\ldots,288,
\end{equation*}
where $N_k$ is the number of observations at period $k$.  Define
$\hat{\beta}=(\hat{\beta}_1,\ldots,\hat{\beta}_{288})'$ and 
$\hat{s}_t=f_t\hat{\beta}$.  The seasonally-adjusted returns are defined as 
$\tilde{y}_t=y_t\hat{s}_t^{-1}$ for normal errors, or
$\tilde{y}_t=\sqrt{\nu/(\nu-2)}y_t\hat{s}_t^{-1}$ for student-t errors.
We then use the adjusted returns to estimate the GARCH models 
above using maximum likelihood methods.

\vspace{.1cm}
\item {\bf Daily AR-RV Models.}
Let $x_t$ denote the daily realized variance.  Following \cite{AndeBollDieb:03}, we 
considered a number of fractional (long-memory) ARMA$(p,q)$ models for $x_t$ and $\log x_t$ 
for different values of $p$ and $q$.  BIC identified the best models as a fractional AR(2)
model for $x_t$, and a fractional AR(1) for $\log x_t$.  The models are 
\begin{eqnarray*}
(1-\phi_1 B -\phi_2 B^2)(1-B)^d x_t &=& \alpha + \varepsilon_t,  ~~~ \varepsilon_t \sim \mathcal{N}(0,1)\\
(1-\phi_1 B)(1-B)^d \log(x_t) &=& \alpha + \varepsilon_t,  ~~~ \varepsilon_t \sim \mathcal{N}(0,1)
\end{eqnarray*}

\vspace{.1cm}
\item {\bf Daily Realized GARCH Models.}
Let $y_t$ denote the daily return, $x_t$ the observed daily realized variance, 
and $h_t$ the unobserved daily variance on day $t$. 
The linear Realized GARCH(1,1) model is
\begin{eqnarray*}
y_t &=& \sqrt{h_t}z_t, ~~~ z_t \sim \mathcal{N}(0,1)\\
x_t &=& \xi + \phi h_t + \tau_1 z_t + \tau_2(z_t^2-1) + \sigma_u u_t, ~~~ u_t \sim \mathcal{N}(0,1)\\
h_t &=& \omega + \beta h_{t-1} + \gamma x_{t-1}.
\end{eqnarray*}
The log-linear Realized GARCH(1,1) model is
\begin{eqnarray*}
y_t &=& \sqrt{h_t}z_t, ~~~ z_t \sim \mathcal{N}(0,1)\\
\log(x_t) &=& \xi + \phi \log(h_t) + \tau_1 z_t + \tau_2(z_t^2-1) + \sigma_u u_t, ~~~ u_t \sim \mathcal{N}(0,1)\\
\log(h_t) &=& \omega + \beta \log(h_{t-1}) + \gamma \log(x_{t-1}).
\end{eqnarray*}

\end{itemize}

\newpage
\input announcements.tex

\vspace{3cm}
~

\end{document}

%% file: model-mnemonics-0.tex
\begin{table}[tbp]
\begin{center}
\begin{tabular}{lccccccc}
  \hline
  \hline
               & SV      & Return & Leverage & Return & Volatility & Seasonal & Announcement \\
     Model     & Factors & Errors &  Effect  & Jumps  &  Jumps     & Effects  & Effects      \\
 \hline
  SV$_i$       & $i$ & Normal    &   &   &    & x & x \\
  ASV$_i$      & $i$ & Normal    & x &   &    & x & x \\
  SVt$_i$      & $i$ & Student-t & x &   &    & x & x \\
  SVJ$_i$      & $i$ & Normal    & x & x &    & x & x \\ 
  SVCJ$_i$     & $i$ & Normal    & x & x & x  & x & x \\
  \hline
   \hline
\end{tabular}
\caption{Mnemonics for the stochastic volatility models that we consider.  Here, $i=$ 1 or 2.} 
\label{Models}
\end{center}
\end{table}

%% file: model-selection-table-2-2-288-long.tex
\begin{table}[t]
\centering
\begin{tabular}{lccccccr}
  \hline \hline 
              & $d^*$ & $d_s$ & $d_a$ & $d$ & $\log \mathcal{L}$ & BIC & $-2\log \mathcal{B}_{ij}$ \\ 
   \hline
  GARCH        & 3 & 279 & 0 & 282 & 192558 & -381775 & 10584 \\ 
  GARCH-t      & 4 & 279 & 0 & 283 & 197725 & -392097 & 262 \\ 
  GJR          & 4 & 279 & 0 & 283 & 192662 & -381969 & 10390 \\ 
  GJR-t        & 5 & 279 & 0 & 284 & 197795 & -392224 & 135 \\ 
  EGARCH       & 4 & 279 & 0 & 283 & 192459 & -381564 & 10795 \\ 
  EGARCH-t     & 5 & 279 & 0 & 284 & 197759 & -392153 & 206 \\ 
  SV$_2$       & 6 & 210 & 51 & 267 & 198788 & -394424 & -2065 \\ 
  ASV$_2$      & 7 & 210 & 51 & 268 & 198865 & -394566 & -2207 \\ 
  SVt$_2$      & 8 & 196 & 51 & 255 & 199235 & -395461 & -3102 \\ 
  SVJ$_2$      & 10 & 191 & 52 & 253 & 198958 & -394929 & -2570 \\ 
  SVCJ$_2$     & 12 & 189 & 52 & 254 & 199214 & -395429 & -3070 \\ 
   \hline
\hline
\end{tabular}
\caption{Degrees of freedom, log-likelihoods, BIC statistics and approximate log Bayes Factors
	 for each model (relative to the SV$_1$ model) for the estimation period, March 2007--March 2009. } 
\label{Modelfit}
\end{table}

%% file: param-estimates-two-factor-stacked-288.tex
\begin{table}[bpt]
\centering
\begin{tabular}{cccccc}
  \hline
\hline
  & SV$_2$ & ASV$_2$ & SVt$_2$ & SVJ$_2$ & SVCJ$_2$ \\ 
  \hline
       $\mu$ & 0.0001 & 0.0000 & 0.0000 & 0.0000 & 0.0000 \\ 
             & {\small (0.0001)} & {\small (0.0001)} & {\small (0.0001)} & {\small (0.0001)} & {\small (0.0001)} \\ 
    $\sigma$ & 0.060 & 0.060 & 0.059 & 0.060 & 0.059 \\ 
             & {\small (0.011)} & {\small (0.012)} & {\small (0.014)} & {\small (0.012)} & {\small (0.013)} \\ 
    $\phi_1$ & 0.9998 & 0.9998 & 0.9999 & 0.9999 & 0.9999 \\ 
             & {\small (0.0001)} & {\small (0.0001)} & {\small (0.0000)} & {\small (0.0001)} & {\small (0.0001)} \\ 
  $\sigma_1$ & 0.022 & 0.021 & 0.019 & 0.020 & 0.020 \\ 
             & {\small (0.001)} & {\small (0.002)} & {\small (0.002)} & {\small (0.001)} & {\small (0.001)} \\ 
    $\tau_1$ & 1.257 & 1.300 & 1.339 & 1.291 & 1.277 \\ 
             & {\small (0.246)} & {\small (0.411)} & {\small (0.452)} & {\small (0.606)} & {\small (0.373)} \\ 
    $\phi_2$ & 0.927 & 0.929 & 0.957 & 0.946 & 0.945 \\ 
             & {\small (0.004)} & {\small (0.004)} & {\small (0.003)} & {\small (0.003)} & {\small (0.004)} \\ 
  $\sigma_2$ & 0.191 & 0.189 & 0.138 & 0.158 & 0.127 \\ 
             & {\small (0.005)} & {\small (0.005)} & {\small (0.005)} & {\small (0.004)} & {\small (0.005)} \\ 
    $\tau_2$ & 0.509 & 0.512 & 0.476 & 0.488 & 0.486 \\ 
             & {\small (0.007)} & {\small (0.009)} & {\small (0.010)} & {\small (0.009)} & {\small (0.011)} \\ 
      $\rho$ &  & -0.095 & -0.126 & -0.106 & -0.136 \\ 
             &  & {\small (0.014)} & {\small (0.017)} & {\small (0.015)} & {\small (0.019)} \\ 
       $\nu$ &  &  & 20.58 &  &  \\ 
             &  &  & {\small (1.12)} &  &  \\ 
    $\kappa$ &  &  &  & 0.0018 & 0.0042 \\ 
             &  &  &  & {\small (0.0003)} & {\small (0.0004)} \\ 
     $\mu_y$ &  &  &  & 0.060 & -0.007 \\ 
             &  &  &  & {\small (0.036)} & {\small (0.013)} \\ 
  $\sigma_y$ &  &  &  & 0.437 & 0.202 \\ 
             &  &  &  & {\small (0.046)} & {\small (0.015)} \\ 
     $\mu_v$ &  &  &  &  & 0.816 \\ 
             &  &  &  &  & {\small (0.086)} \\ 
  $\sigma_v$ &  &  &  &  & 1.220 \\ 
             &  &  &  &  & {\small (0.069)} \\ 
   \hline
  aprob & 0.308 & 0.312 & 0.328 & 0.289 & 0.292 \\ 
  ineff & 51.5 & 41.0 & 90.8 & 29.0 & 97.9 \\ 
   \hline
\hline
\end{tabular}
\caption{Posterior means and standard deviations (in parentheses) for the two-factor models.
The bottom two rows are the Metropolis-Hastings acceptance probabilities and inefficiency factors 
for the slowest mixing parameter, $\sigma_1$.}
\label{parameters}
\end{table}

%% file: vol-decomp-2007-2009-all-1-1-0-2-2-288.tex
\begin{table}[bpt]
\begin{center}
\begin{tabular}{lcccccccccc}
  \hline \hline && \multicolumn{4}{c}{Log Variance} && \multicolumn{4}{c}{Volatility} \\ 
                 \cline{3-6} \cline{8-11} 
 
                 && $x_1$ & $x_2$ & $s$ & $a$ && $X_1$ & $X_2$ & $S$ & $A$ \\ 
 \hline
  SV$_2$ &  & 53.4 & 7.0 & 38.3 & 1.3 &  & 59.1 & 7.2 & 30.4 & 3.3 \\ 
  ASV$_2$ &  & 53.5 & 6.8 & 38.4 & 1.3 &  & 59.1 & 7.1 & 30.5 & 3.3 \\ 
  SVt$_2$ &  & 53.5 & 6.4 & 38.8 & 1.3 &  & 58.9 & 7.1 & 30.8 & 3.3 \\ 
  SVJ$_2$ &  & 53.3 & 6.9 & 38.6 & 1.3 &  & 58.9 & 7.4 & 30.6 & 3.1 \\ 
  SVCJ$_2$ &  & 52.1 & 8.6 & 38.0 & 1.2 &  & 57.0 & 10.2 & 29.7 & 3.2 \\ 
   \hline
\hline
\end{tabular}
\caption{Volatility decomposition (percentage of total), March 2007--March 2009}
\label{vardecomps}
\end{center}
\end{table}

%% file: test-rv.tex
\begin{table}[bpt]
\centering
\begin{tabular}{lcccccccc}
  \hline \hline 
            &&  \multicolumn{3}{c}{1 Hour} &&  \multicolumn{3}{c}{Daily} \\
           \cline{3-5} \cline{7-9} 
             && Bias & MAE & $R^2$ && Bias & MAE & $R^2$ \\ 
   \hline
   EWMA          &  & -0.012 & 0.068 & 59.0 &  & 0.129 & 0.314 & 49.7 \\ 
   GARCH         &  & -0.024 & 0.073 & 56.8 &  & -0.177 & 0.321 & 49.2 \\ 
   GARCH-t       &  & -0.016 & 0.071 & 56.4 &  & -0.169 & 0.320 & 47.2 \\ 
   GJR        &  & -0.023 & 0.072 & 57.4 &  & -0.166 & 0.316 & 48.8 \\ 
   GJR-t      &  & -0.016 & 0.070 & 56.9 &  & -0.156 & 0.316 & 46.8 \\ 
   EGARCH        &  & -0.028 & 0.072 & 61.3 &  & -1.132 & 1.146 & 57.5 \\ 
   EGARCH-t      &  & -0.018 & 0.068 & 60.6 &  & -0.491 & 0.543 & 57.4 \\ 
   SV$_2$        &  & -0.006 & 0.061 & 66.2 &  & -0.048 & 0.201 & 73.5 \\ 
   ASV$_2$       &  & -0.005 & 0.060 & 66.4 &  & -0.050 & 0.205 & 72.4 \\ 
   SVt$_2$       &  & -0.004 & 0.060 & 66.5 &  & -0.053 & 0.204 & 72.1 \\ 
   SVJ$_2$       &  & -0.007 & 0.060 & 66.4 &  & -0.077 & 0.212 & 72.7 \\ 
   SVCJ$_2$      &  & -0.007 & 0.060 & 66.1 &  & -0.090 & 0.216 & 73.5 \\ 
   AR-RV$^*$     &  &  &  &  &  & -0.090 & 0.229 & 69.0 \\ 
   AR-RV$^\dagger$ &  &  &  &  &  & 0.013 & 0.208 & 67.5 \\ 
   RealGARCH$^*$   &  &  &  &  &  & -0.571 & 0.616 & 61.6 \\ 
   RealGARCH$^\dagger$ &  &  &  &  &  & -0.082 & 0.226 & 68.7 \\ 
   \hline
\hline
\end{tabular}
\caption{Out-of-sample predictions for hourly and daily realized volatility.
The AR-RV and RealGARCH models are estimated using daily data.  
$*$ and $\dagger$ indicate models fit on the linear and log-linear scale, respectively.
All other models are estimated using 5-minute data.} 
\label{RVdaily}
\end{table}

%% file: test-reg.tex
\begin{table}[bpt]
\centering
\begin{tabular}{lcccccc}
  \hline \hline 
           && \multicolumn{2}{c}{1 Hour} &&  \multicolumn{2}{c}{Daily} \\
           \cline{3-4} \cline{6-7} 
                 && $b_1$ ($t$) & $b_2$ ($t$) && $b_1$ ($t$) & $b_2$ ($t$) \\ 
   \hline
   EWMA          &&  0.00 ( 0.09) & 0.91 (61.84) &&  0.06 ( 2.12) & 0.90 (26.55) \\ 
   GARCH         && -0.02 (-1.74) & 0.93 (70.85) &&  0.08 ( 2.35) & 0.90 (26.91) \\ 
   GARCH-t       && -0.02 (-1.69) & 0.93 (72.58) &&  0.07 ( 2.06) & 0.91 (27.90) \\ 
   GJR           && -0.00 (-0.34) & 0.91 (68.56) &&  0.10 ( 3.08) & 0.88 (27.27) \\ 
   GJR-t         && -0.00 (-0.28) & 0.91 (70.31) &&  0.09 ( 2.73) & 0.89 (28.23) \\ 
   EGARCH        && -0.01 (-0.83) & 0.92 (51.04) &&  0.11 ( 2.88) & 0.86 (21.94) \\ 
   EGARCH-t      && -0.00 (-0.30) & 0.92 (54.43) &&  0.15 ( 3.08) & 0.86 (22.06) \\ 
   AR-RV$^*$           &&         &              &&  0.19 ( 2.75) & 0.79 (11.91) \\ 
   AR-RV$^\dagger$     &&         &              &&  0.08 ( 1.09) & 0.89 (13.39) \\ 
   RealGARCH$^*$       &&         &              && -0.03 (-0.36) & 0.98 (18.68) \\ 
   RealGARCH$^\dagger$ &&         &              &&  0.04 ( 0.60) & 0.91 (11.88) \\ 
   \hline\hline
\end{tabular}
\caption{Bivariate 'horse-race' regressions for realized volatility using the model in Equation 
(\ref{eqn:horse-race}).  $b$ and $t$ represent the estimated regression coefficients and 
corresponding $t$ statistics.  AR-RV and RealGARCH models are estimated using daily data.  
$*$ and $\dagger$  indicate models fit on the linear and log-linear scale, respectively.
All other models are estimated using 5-minute data.} 

\label{RV-multiple}
\end{table}

%% file: test-ret1.tex
\begin{table}[bpt]
\centering
\begin{tabular}{lccccccccccccccc}
  \hline \hline 
            &&  \multicolumn{4}{c}{5-Minute} 
            &&  \multicolumn{4}{c}{1 Hour}   
            &&  \multicolumn{4}{c}{Daily} \\
           \cline{3-6} \cline{8-11} \cline{13-16} 
             && 1 & 5 & 10 & $D$ 
             && 1 & 5 & 10 & $D$ 
             && 1 & 5 & 10 & $D$ \\ 
   \hline
   GARCH         &  & 1.6 & 4.6 & 8.3 & 25 &  & 1.3 & 4.0 & 7.4 & 30 &  & 0.8 & 5.9 &  9.9 & 36 \\ 
   GARCH-t       &  & 1.6 & 4.7 & 8.5 & 24 &  & 1.2 & 4.2 & 8.0 & 24 &  & 0.6 & 6.7 & 11.1 & 36 \\ 
   GJR           &  & 1.6 & 4.7 & 8.4 & 25 &  & 1.2 & 4.0 & 7.5 & 30 &  & 0.6 & 5.8 & 10.0 & 28 \\ 
   GJR-t         &  & 1.6 & 4.7 & 8.5 & 24 &  & 1.2 & 4.2 & 8.1 & 23 &  & 0.8 & 6.4 & 11.1 & 28 \\ 
   EGARCH        &  & 1.5 & 4.4 & 8.0 & 29 &  & 1.1 & 3.5 & 6.9 & 33 &  & 0.0 & 0.5 &  2.1 & 96 \\ 
   EGARCH-t      &  & 1.0 & 4.9 & 9.8 & 13 &  & 1.0 & 3.8 & 7.5 & 26 &  & 0.0 & 2.4 &  6.6 & 47 \\ 
   SV$_2$        &  & 1.3 & 5.0 & 9.4 & 14 &  & 1.3 & 4.4 & 8.7 & 17 &  & 2.1 & 6.4 &  9.8 & 28 \\ 
   ASV$_2$       &  & 1.3 & 5.0 & 9.4 & 15 &  & 1.2 & 4.3 & 8.6 & 17 &  & 1.7 & 6.0 &  9.5 & 34 \\ 
   SVt$_2$       &  & 1.3 & 5.0 & 9.6 & 14 &  & 1.3 & 4.3 & 8.6 & 17 &  & 1.8 & 6.4 &  9.6 & 31 \\ 
   SVJ$_2$       &  & 1.4 & 5.1 & 9.5 & 14 &  & 1.1 & 4.2 & 8.5 & 18 &  & 1.7 & 5.9 &  9.5 & 31 \\ 
   SVCJ$_2$      &  & 1.4 & 5.1 & 9.5 & 14 &  & 1.1 & 4.2 & 8.6 & 17 &  & 1.2 & 5.7 &  9.5 & 32 \\ 
   RealGARCH$^*$ &  &  &  &  &  &  &  &  &  &  &  & 1.4 & 4.9 & 8.2 & 41 \\ 
   RealGARCH$^\dagger$ &  &  &  &  &  &  &  &  &  &  &  & 1.7 & 5.0 & 8.1 & 43 \\ 
   \hline
\hline
\end{tabular}
\caption{Out-of-sample lower-tail coverage probabilities (1\%, 5\% and 10\%) and 
distance metrics ($D$) for 5-minute, hourly and daily returns.  All values are multiplied
by 100.  RealGARCH models are estimated using daily data.  $*$ and $\dagger$ indicate models fit on the linear 
and log-linear scale, respectively.  All other models are estimated using 
5-minute returns.} 
\label{Returns}
\end{table}

%% file: vxx-trading-table.tex
\begin{table}[t]
\begin{center}
\begin{tabular}{lrrrrrrrrrrrr}
  \hline \hline 
              && \multicolumn{5}{c}{Single Model} && \multicolumn{5}{c}{Long Minus Short} \\
                  \cline{3-7}  \cline{9-13} 
              && $+$ & $-$ & Mean & SD & SR && $+$ & $-$ & Mean & SD & SR \\ 
   \hline
   GARCH            &  &   3 & 129 & 25.4 & 26.6 & 0.95 &  &  62 & 325 & 67.8 & 40.3 & 1.68 \\ 
   GARCH-t          &  &   1 &  62 & 16.0 & 15.1 & 1.06 &  &   7 & 335 & 77.2 & 37.4 & 2.06 \\ 
   GJR              &  &   4 & 154 & 29.6 & 28.8 & 1.03 &  &  61 & 300 & 63.6 & 38.8 & 1.64 \\ 
   GJR-t            &  &   1 &  62 & 16.0 & 15.1 & 1.06 &  &   7 & 335 & 77.2 & 37.4 & 2.06 \\ 
   EGARCH           &  & 450 &   2 &-40.0 & 44.0 &-0.91 &  &   0 & 405 &133.2 & 70.6 & 1.89 \\ 
   EGARCH-t         &  &   2 &  14 &  4.8 &  5.9 & 0.81 &  &   7 & 384 & 88.4 & 39.9 & 2.22 \\ 
   SV$_2$           &  &   7 & 551 &105.1 & 48.0 & 2.19 &  & 189 &  34 &-11.9 & 29.4 &-0.41 \\ 
   ASV$_2$          &  &   7 & 532 & 94.5 & 47.0 & 2.01 &  & 159 &  23 & -1.3 & 25.8 &-0.05 \\ 
   SVt$_2$          &  &   5 & 521 & 82.6 & 46.5 & 1.78 &  & 140 &  13 & 10.6 & 24.9 & 0.43 \\ 
   SVJ$_2$          &  &   7 & 431 & 90.4 & 44.5 & 2.03 &  &  52 &  17 &  2.8 & 20.7 & 0.14 \\ 
   SVCJ$_2$         &  &   7 & 396 & 93.2 & 40.2 & 2.32 &  &   &  &   &   &  \\ 
   AR-RV$^*$        &  &   8 &   0 &  3.2 & 10.2 & 0.32 &  &   0 & 397 & 90.0 & 39.5 & 2.28 \\ 
   AR-RV$^\dagger$     & & 9 & 395 & 53.2 & 42.9 & 1.24 &  & 123 & 124 & 40.0 & 36.9 & 1.08 \\ 
   RealGARCH$^*$       & & 6 &   0 &  4.7 & 11.8 & 0.40 &  &   3 & 398 & 88.5 & 41.5 & 2.13 \\ 
   RealGARCH$^\dagger$ & & 3 & 461 & 73.9 & 44.4 & 1.66 &  & 128 &  59 & 19.3 & 30.2 & 0.64 \\ 
   \hline
\hline
\end{tabular}
\caption{Out-of-sample VXX trading results, March 2009--March 2012.   
Trading rules are based on a single model (left panel), and a long-minus-short portfolio
(right panel).
$+$ and $-$ are the number of days each portfolio is long or short the index.
Mean, SD and SR are the mean, standard deviation and Sharpe ratio (Mean/SD)
for the portfolio returns.}
\end{center}
\end{table}

%% file: announcements.tex
\begin{table}[tbp] 
\begin{center}
\begin{tabular}{clllcc}
\hline
\hline
No.& Announcement Type          & Frequency  &  Days    &  Time \\
\hline
 1 & ADP Employment             & Monthly    & Wed-Thu  &  8:15 \\
 2 & Jobless Claims             & Weekly     & Thu      &  8:30 \\
 3 & Consumer Price Index       & Monthly    & Wed-Fri  &  8:30 \\
 4 & Durable Goods              & Monthly    & Wed-Fri  &  8:30 \\
 5 & GDP Advance                & Quarterly  & Thu-Fri  &  8:30 \\
 6 & Monthly Payrolls           & Monthly    & Fri      &  8:30 \\
 7 & Empire State Manuf.        & Monthly    & Mon-Fri  &  8:30 \\
 8 & Consumer Confidence        & Monthly    & Tue-Wed  & 10:00 \\
 9 & Philadelphia Fed           & Monthly    & Thu      & 10:00 \\
10 & ISM Manufacturing          & Monthly    & Mon-Fri  & 10:00 \\
11 & ISM Services               & Monthly    & Tue-Fri  & 10:00 \\
12 & FOMC Minutes               & 8/year     & Tue-Wed  & 14:00 \\
13 & FOMC                       & 8/year     & Tue-Wed  & 14:15 \\
14 & Sunday Open                & Weekly     & Sun      & 18:00 \\
\hline
\hline
\end{tabular}
\caption{List of major US macroeconomic announcements that we incorporate in our analysis,
and the frequency, days of the week, and time of day (ET) when the announcements occur.}
\label{announcements}
\end{center}
\end{table}